\documentclass{article}

\usepackage{microtype}
\usepackage{graphicx}
\usepackage{subfigure}
\usepackage{booktabs}
\usepackage{xcolor}

\usepackage{float}
\usepackage[hypertexnames=false]{hyperref}
\usepackage{xurl} 

\usepackage[accepted]{mlsys2025}
\usepackage[pass]{geometry}
\makeatletter
\renewcommand{\Notice@String}{}
\makeatother
\hypersetup{pdfauthor={Shuo Rong, Ruibo Fan, Jiawen Wen, Xiaowen Chu},pdfsubject={}}
\usepackage{array}
\usepackage{placeins}
\usepackage{amsmath,amssymb}

\let\algorithmiccomment\relax

\usepackage{algpseudocode}
\usepackage{multirow}
\usepackage{pifont}

\providecommand{\Description}[1]{}

\newcommand{\vllmver}{0.26.0}

\definecolor{algtype}{RGB}{0,80,180}
\definecolor{algkw}{RGB}{140,40,160}
\definecolor{algcomment}{RGB}{190,90,20}
\definecolor{algnote}{RGB}{0,120,130}
\newcommand{\Type}[1]{\textcolor{algtype}{\textbf{#1}}}
\newcommand{\Kw}[1]{\textcolor{algkw}{\textbf{#1}}}
\newcommand{\AlgComment}[1]{\Statex \textcolor{algcomment}{\footnotesize // #1}}

\algrenewcommand\algorithmiccomment[1]{\hfill$\triangleright$ #1}

\newcommand{\resserverunningtitle}{ResServe: Accurate and Efficient FP4 LLM Inference}
\mlsystitlerunning{\resserverunningtitle}

\begin{document}

\newgeometry{margin=1in,ignoreheadfoot,headheight=12pt,headsep=12pt,footskip=24pt,columnsep=0.25in}
\setlength{\headwidth}{\textwidth}

\twocolumn[
\mlsystitle{ResServe: Accurate and Efficient FP4 LLM Inference with Residue Compensation}

{\begin{center}
{\large
\textbf{Shuo Rong}\textsuperscript{1}\hspace{1.5em}
\textbf{Ruibo Fan}\textsuperscript{1}\hspace{1.5em}
\textbf{Jiawen Wen}\textsuperscript{1}\hspace{1.5em}
\textbf{Xiaowen Chu}\textsuperscript{1,2}\par}
\vspace{6pt}
{\small
\textsuperscript{1}The Hong Kong University of Science and Technology (Guangzhou), Guangzhou, China\\
\textsuperscript{2}The Hong Kong University of Science and Technology, Hong Kong SAR, China\\[4pt]
\texttt{\{srongaa, ruibo.fan, jwen341\}@connect.hkust-gz.edu.cn}\\
\texttt{xwchu@hkust-gz.edu.cn}}
\end{center}}
\mlsyskeywords{LLM inference, Quantization}

\vskip 0.12in
\begin{abstract}
Block-scaled 4-bit floating-point (FP4) quantization reduces large language model inference costs but can degrade task quality.
Higher-precision computation can mitigate this loss at the expense of some of the efficiency gains from FP4 inference.
Narrow matrix computations common in decoding often leave spare hardware capacity for low-cost precision recovery.
ResServe combines an FP4 activation base with an FP4-quantized residue representing information lost during quantization.
It uses this spare capacity to compute the residue's contribution and combines results on-chip, avoiding extra result transfers.
We apply this principle to linear layers and decode attention, while selective compensation limits added work when processing input prompts.
On NVIDIA B200, compensated linear kernels using the NVFP4 format improve numerical precision while retaining latency close to uncompensated execution for the evaluated operations with few rows.
In the evaluated NVFP4 inference configurations, ResServe adds at most 3.4\% request-latency overhead over W4A4 and reduces request latency by 22.4--53.3\% relative to W4A8.
Across five models, code-generation accuracy improves over same-format uncompensated inference, with mean relative gains of 15.04\% for NVFP4 and 53.57\% for MXFP4.
\end{abstract}
]

\fancyhead[C]{\small\bfseries\resserverunningtitle}
\thispagestyle{plain}

\noindent Our code is available at \href{https://github.com/ShawRong/resserve}{github.com/ShawRong/resserve}.

\section{Introduction}
\label{sec:intro}
Large language models (LLMs) support coding assistants~\cite{chen_evaluating_2021}, agentic tool use~\cite{schick_toolformer_2023}, and multi-turn chat.
Reducing inference costs while preserving model quality is important for these applications.
Quantization reduces these costs by lowering the precision of model weights and activations, the intermediate values produced during computation~\cite{frantar_optq_2023,lin_awq_2024,xiao_smoothquant_2023}.
Recent GPUs natively support four-bit floating-point (FP4) computation~\cite{nvidiacorporation_blackwell_2026a}, offering lower storage requirements and lower latency.
However, limited precision introduces quantization error that can degrade model quality~\cite{egiazarian_bridging_2025,cook_four_2026}.

This matters especially for complex reasoning tasks.
Models such as DeepSeek-R1 and Qwen3 generate reasoning traces to solve mathematical and coding problems~\cite{guo_deepseekr1_2025,yang_qwen3_2025}.
Low-bit quantization can reduce reasoning accuracy~\cite{liu_quantization_hurts_2025}.
Under some quantization settings, models also generate longer chains of thought while producing less accurate answers~\cite{lotfi_quantized_reasoning_2026}.
Generating more tokens can offset some of the inference speedup from quantization.
Precision recovery may therefore affect both answer quality and the computation needed to produce an answer.

Prior work mitigates low-bit quantization error through improved quantization algorithms and transformations of weights and activations~\cite{frantar_optq_2023,lin_awq_2024,dettmers_spqr_2023,kim_squeezellm_2024,xiao_smoothquant_2023,shao_omniquant_2024,ashkboos_quarot_2024,liu_spinquant_2025,sun_flatquant_2025,egiazarian_bridging_2025}.
Unlike weights, which can be quantized offline, activations must be quantized during inference.
Online quantization and associated transformations therefore require efficient execution~\cite{xiao_smoothquant_2023,ashkboos_quarot_2024,egiazarian_bridging_2025}.
Another approach pairs four-bit weights with eight-bit activations~\cite{lin_qserve_2025,zheng_mixllm_2024,zhang_lqer_2024,baek_fireq_2025}.
These configurations can improve activation precision, but their execution paths do not retain all the computational advantages of native FP4~\cite{jarmusch_dissecting_2025,nvidiacorporation_blackwell_2026a}.
This raises the question: \emph{how can we recover precision in FP4 inference while retaining its execution efficiency?}

Our observation is that additional computation does not proportionally increase latency.
During decoding, which generates tokens one step at a time, small batches limit the number of token rows processed by linear layers.
Decode attention processes few queries for the current token against keys accumulated over the context.
The resulting matrix computations are narrow and may leave GPU resources underused.
This spare capacity offers room for additional precision-recovery computation.
Exploiting it requires suitable execution layouts and low-cost handling of intermediate results.

ResServe uses this spare capacity to compute a correction for activation quantization error.
It quantizes the difference between the original activation and its FP4 approximation into FP4.
ResServe computes both the base output and a correction from the quantized residue, combining them to recover precision.
We call this process residue compensation.
ResServe adapts how compensation is computed to the matrix dimensions and combines base and compensation results on-chip, avoiding separate writes to GPU memory followed by an addition.
During decoding, it applies this principle to linear layers and attention.
Prefill processes the input prompt and typically handles more tokens at once, making full compensation more costly.
ResServe compensates a subset of activation channels during prefill to control the added computation, a strategy we call selective prefill compensation.

We implement ResServe in vLLM~\cite{kwon_efficient_2023}, integrating compensation for linear layers and decode attention.
The system supports NVFP4 and MXFP4 for linear computation and uses an NVFP4 KV cache for attention.
Relative to same-format uncompensated inference, code-generation accuracy improves by an average of 15.04\% for NVFP4 and 53.57\% for MXFP4 across five models.
With MXFP4, mean relative accuracy gains on the HMMT and AIME mathematics benchmarks~\cite{dekoninck2026matharena} are 50.23\% and 33.34\%, respectively.
On the same code-generation benchmark, total output tokens decrease by an average of 3.18\% for NVFP4 and 9.16\% for MXFP4.

ResServe retains near-W4A4 request latency in the evaluated NVFP4 configurations with selective prefill disabled, while reducing latency relative to W4A8 by 22.4\% on Qwen3-8B and 53.3\% on Qwen3-32B.
With the prefill compensation ratio of 0.25 used in the quality evaluation, MXFP4 ResServe adds 7.0--9.4\% request-latency overhead over same-format W4A4.

This paper makes the following contributions:
\begin{list}{\labelitemi}{%
    \setlength{\leftmargin}{\leftmargini}%
    \setlength{\topsep}{4pt}%
    \setlength{\partopsep}{1pt}%
    \setlength{\itemsep}{2pt}%
    \setlength{\parsep}{2pt}}
    \item We identify spare hardware capacity in narrow decoding operators that can accommodate additional low-precision computation for precision recovery without a proportional latency increase.
    \item We design execution layouts that use this spare capacity for residue compensation in linear layers and decode attention.
    On-chip result combination and selective prefill compensation control the added execution cost.
    \item We build a complete inference system with NVFP4/MXFP4 linear computation and NVFP4 KV caching.
    The system improves mean coding and mathematics accuracy across five models, while low-load NVFP4 serving measurements show low integration overhead.
\end{list}

\section{Background}
\label{sec:background}

\label{sec:nvfp4_representation}

Block-scaled FP4 represents tensor elements with four-bit floating-point values and scales shared within groups.
NVFP4 and microscaling FP4 (MXFP4) both use E2M1 values but differ in scaling.
NVFP4 shares an FP8 E4M3 scale within each 16-element block and an additional 32-bit floating-point (FP32) scale across the tensor~\cite{nvidia_pretraining_2026,cook_four_2026}.
MXFP4 shares an E8M0 power-of-two scale across 32 consecutive elements, without an additional tensor-level scale~\cite{rouhani_ocp_2023}.
Figure~\ref{fig:nvfp4_background} shows both representations and their linear-layer execution flow.

Here, W4A4 denotes block-scaled FP4 weights and activations.
W4A8 denotes block-scaled FP4 weights paired with FP8 or block-scaled MXFP8 activations.

\begingroup
\setlength{\intextsep}{12pt}
\begin{figure}[!htbp]
    \centering
    \includegraphics[width=0.85\linewidth,trim=1.9bp 2.8bp 1.9bp 2.9bp,clip]{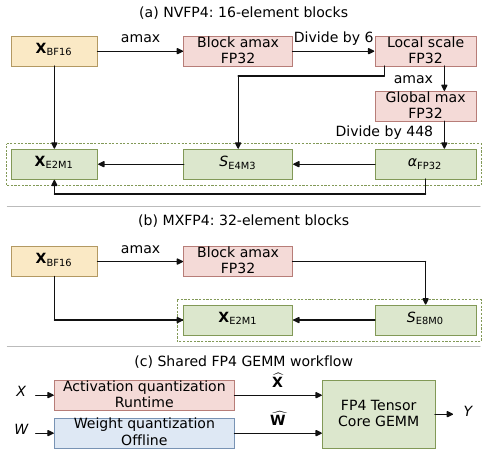}
    \caption{
        NVFP4 and MXFP4 representations and their W4A4 linear-layer execution.
        Activations are quantized at runtime and weights offline.
    }
    \Description{NVFP4 computes 16-element local scales, an FP32 tensor scale, and E4M3 block scales; MXFP4 uses 32-element blocks with E8M0 power-of-two scales. Both produce E2M1 values. The shared workflow feeds dynamically quantized activations and offline-quantized weights to an FP4 Tensor Core GEMM.}
    \label{fig:nvfp4_background}
\end{figure}
\endgroup

We use QDQ to denote quantization followed by dequantization.
For an input $x$, we write the resulting value as $\hat{x}=\mathrm{QDQ}(x)$.
The quantization residue is the difference from the original value, $r=x-\hat{x}$.
The same notation applies to activation, weight, and quantized residue matrices, denoting the numerical values represented by their FP4 encodings rather than the codes themselves.

We write a linear projection as $\mathbf{X}^{M\times K}\mathbf{W}^{K\times N}\rightarrow\mathbf{Y}^{M\times N}$, where $M$ is the actual number of token rows in the current GEMM.
During standard autoregressive decoding, each active request contributes one token row per step.
The dimensions $K$ and $N$ denote the input and output channels, respectively.
An input channel is a column of $\mathbf{X}$ and the corresponding row of $\mathbf{W}$.

\section{Motivation}
\label{sec:motivation}

Precision recovery must account for both quantization error and execution cost.
Plain block-scaled FP4 can leave a task-quality gap relative to BF16 (Table~\ref{tab:quality-system}).
Wider activations can reduce error but sacrifice native FP4's theoretical throughput advantage.
Residue compensation recovers discarded activation information through additional FP4 computation.

Compensation latency depends on how this additional work executes.
A separate compensation GEMM adds a kernel launch and output traffic for storing and adding the correction.
Decode linear layers and attention have narrow token and query dimensions, respectively, which can leave tile capacity underused.
When the original and extended computations fit within the same tile, residue computation can occupy otherwise idle positions.
This leaves the tile count along that dimension unchanged.
Exploiting this capacity requires organizing the compensation computation and merging its results on-chip.
Prefill processes more tokens at once, so compensating only a subset of channels helps control the added computation.

\section{ResServe Design}
\label{sec:design}

\subsection{Overview}
\label{sec:overview}

\begin{figure}[t]
    \centering
    \includegraphics[width=0.90\linewidth,trim=1bp 2bp 1bp 2bp,clip]{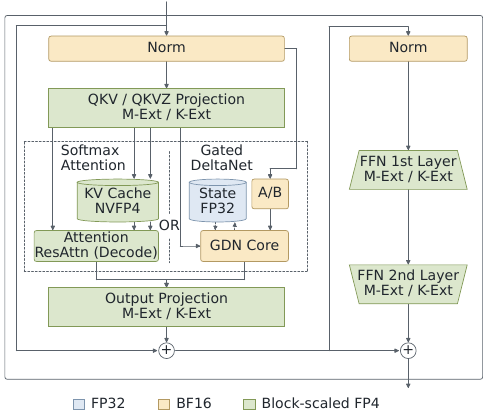}
    \caption{Precision assignment in ResServe. Colors indicate the numerical precision used by each model component.}
    \Description{Block-scaled FP4 is used for QKV or QKVZ, output, and feed-forward projections. The softmax-attention branch uses an NVFP4 KV cache and ResAttn for decode. For models with hybrid attention, Gated DeltaNet retains BF16 core computation and FP32 recurrent state; normalization and A/B computation remain BF16.}
    \label{fig:precision_overview}
\end{figure}

ResServe uses block-scaled FP4 for attention projections and feed-forward linear layers (Figure~\ref{fig:precision_overview}).
ResAttn performs decode attention over the existing paged NVFP4 KV cache.
For models with hybrid attention, Gated DeltaNet retains BF16 core computation and FP32 recurrent state.

\begin{figure}[t]
    \centering
    \includegraphics[width=0.95\linewidth,trim=0.9bp 1.4bp 3.6bp 0.9bp,clip]{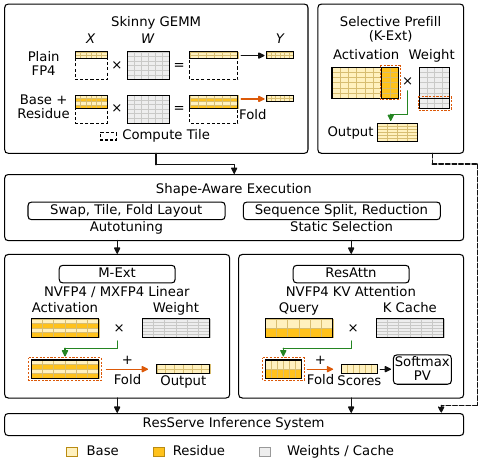}
    \caption{
        ResServe overview. $X$, $W$, and $Y$ denote activations, weights, and outputs.
    }
    \Description{Skinny GEMM tile capacity motivates base--residue compensation. M-Ext folds linear outputs, while ResAttn folds attention scores before softmax. Independent selective prefill support joins these paths in the ResServe inference system.}
    \label{fig:resserve_overview}
\end{figure}

ResServe recovers information lost during activation quantization through additional low-precision computation (Figure~\ref{fig:resserve_overview}).
It computes with an FP4 base and an FP4-quantized residue, combining their contributions on-chip.
This avoids a separate correction tensor in global memory and an extra kernel to add it.

ResServe adapts compensation to the matrix dimensions and to where the base and compensation results must be combined.
For skinny operators common in decoding, M-Ext and ResAttn extend computation along the narrow dimension to use otherwise idle execution capacity.
M-Ext arranges the base and residue as paired activation rows in linear layers, then adds their corresponding outputs.
ResAttn applies the same principle to multiplication of queries and keys, pairing the query base and residue and combining their scores before softmax while preserving the existing NVFP4 KV-cache layout.
For prefill with larger row counts, ResServe selectively compensates residue channels to control the added computation (Section~\ref{sec:prefill_compensation}).

\subsection{On-Chip Residue Compensation}
\label{sec:design_choices}
ResServe adds an activation residue term to the base low-precision linear output:
\begin{equation}
    \label{eq:residue_compensation}
    \mathbf{Y} \;\approx\; \hat{\mathbf{X}}\hat{\mathbf{W}} \;+\; \hat{\mathbf{R}}_{\mathcal{S}}\hat{\mathbf{W}}_{\mathcal{S}},
\end{equation}
where the first term is the plain block-scaled FP4 GEMM and the second recovers information lost to activation quantization.
The residue is $\mathbf{R}=\mathbf{X}-\hat{\mathbf{X}}$, and $\mathcal{S}$ specifies the compensated channels.
The correction uses FP4-quantized values of these residue channels and their corresponding weight rows.
Full compensation includes all channels and reuses the base weight.
Selective compensation includes only a subset to reduce the added computation.
ResServe uses this option for prefill with more token rows; Section~\ref{sec:prefill_compensation} describes channel selection and execution.

\subsection{Full-Residue Decode Compensation (M-Ext)}
\label{sec:shape_adaptive}
\label{sec:small_m_mext}
\label{sec:mext_formulation}
M-Ext organizes full-residue compensation as a GEMM extended along the token dimension.
The base activations and residues multiply the same weight, so M-Ext stacks them as paired rows and adds the corresponding outputs:
\begin{equation}
    \label{eq:mext_fold}
    \mathbf{Z}
    =
    \begin{bmatrix}
    \hat{\mathbf{X}} \\
    \hat{\mathbf{R}}
    \end{bmatrix}
    \hat{\mathbf{W}}
    =
    \begin{bmatrix}
    \hat{\mathbf{X}}\hat{\mathbf{W}} \\
    \hat{\mathbf{R}}\hat{\mathbf{W}}
    \end{bmatrix},
    \qquad
    \mathbf{Y}_{j,:}
    =
    \mathbf{Z}_{j,:} + \mathbf{Z}_{M+j,:}.
\end{equation}
We call the elementwise addition of paired base and residue output rows Fold.
It restores the original $M\times N$ output shape.
This organization doubles the logical row count from $M$ to $2M$ without extending the reduction dimension $K$ or duplicating the weight.
For skinny GEMMs in decoding, the added rows can use otherwise idle execution capacity.
Section~\ref{sec:mext_codesign} develops the layouts and execution.

\subsection{Residue-Compensated Decode Attention (ResAttn)}
\label{sec:resattn}
The multiplication of queries and keys in decode attention is also a skinny GEMM.
In grouped-query attention~\cite{ainslie_gqa_2023}, a small number of queries within each group sharing one KV head is multiplied against keys from the cached sequence.
ResAttn applies the same compensation principle by splitting the BF16 query into an FP4 base and a quantized residue.

Unlike the linear output fold, attention must combine the compensated scores before softmax.
For the query-head group sharing one KV head, let $\hat{\mathbf{Q}}$ and $\hat{\mathbf{R}}$ denote the FP4 base and quantized residue of the BF16 queries.
Each row corresponds to one query head.
The logical computation is
\begin{equation}
    \mathbf{S} \approx
    \frac{\hat{\mathbf{Q}}\hat{\mathbf{K}}^T + \hat{\mathbf{R}}\hat{\mathbf{K}}^T}{\sqrt{d}},\quad
    \mathbf{O} = \mathrm{softmax}(\mathbf{S})\hat{\mathbf{V}}.
\end{equation}
Here $d$ is the head dimension.

ResAttn preserves the existing paged NVFP4 KV-cache layout.
Section~\ref{sec:resattn_execution} explains the matrix multiplications under this layout, together with sequence splitting and reduction selection.

\section{Efficient Execution and System Integration}
\label{sec:runtime_execution}
\label{sec:fused_quantize_extend}

ResServe fuses residue generation and quantization into the activation quantization pass.
The quantizer forms residues from the input and reconstructed base values in registers, avoiding a separate pass to read the base operand.
It directly emits the layout consumed by the FP4 GEMM, eliminating a separate activation reordering kernel.
Appendix~\ref{app:fused_quantization} details the NVFP4 scale computation and fused quantization procedure.

\subsection{M-Ext Kernel: Tiling and Output Folding}
\label{sec:mext_codesign}
\label{sec:tactic_selection}
M-Ext combines operand swapping, tile selection, and output layouts to use spare execution capacity and implement Fold in Eq.~\ref{eq:mext_fold} efficiently.
The layouts place each base output and its corresponding residue output in the same thread's registers, where they are added.

\textbf{Swapping the operands maps tokens to the narrower dimension of the matrix multiply-accumulate (MMA) output tile.} A conventional linear layer evaluates $\hat{\mathbf{X}}^{M\times K}\hat{\mathbf{W}}^{K\times N}$, producing $N$ output features (channels) for each of $M$ tokens.
In the NVFP4 kernels used here, MMA output tiles have a larger minimum row count than column count~\cite{nvidiacorporation_blackwell_2026a,ye_flashinfer_2025a}.
The conventional layout assigns tokens to output rows, so small decode batches leave many tile rows unused.

M-Ext applies this transformation to its stacked base--residue operand $\hat{\mathbf{A}}=[\hat{\mathbf{X}};\hat{\mathbf{R}}]\in\mathbb{R}^{2M\times K}$:
\begin{equation}
    \label{eq:mext_axis_swap}
    \hat{\mathbf{W}}^{T,N\times K}
    \hat{\mathbf{A}}^{T,K\times 2M}
    =
    \mathbf{Z}^{T,N\times 2M}.
\end{equation}
After the swap, rows of the result tile correspond to output channels, and columns correspond to the base and residue rows.
For small decode batches, the kernel can therefore use tiles with fewer columns.
Let $B$ denote the number of activation rows processed by the GEMM, including residue rows when compensation is enabled.
Plain NVFP4 has $B=M$, whereas M-Ext has $B=2M$.

\begin{figure}[!tb]
    \centering
    \includegraphics[width=\linewidth]{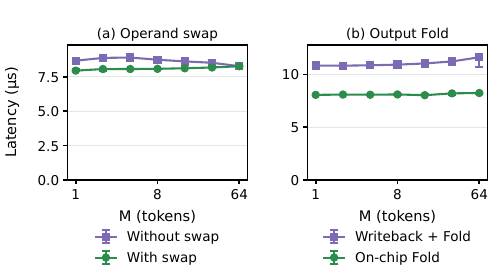}
    \caption{NVFP4 execution ablations. (a) GEMM with and without operand swapping. (b) On-chip Fold versus a separate Fold kernel.
    Timing excludes quantization; bars show min--max across three processes.}
    \label{fig:mext_execution_ablation}
\end{figure}

Figure~\ref{fig:mext_execution_ablation}a compares both orientations for plain NVFP4 GEMM with $N=K=4096$ and a fixed $128\times64$ tile.
Swapping gives $1.04$--$1.10\times$ speedups for $M=1$--$32$, while the two orientations are nearly equal at $M=64$.

\begin{figure}[!tb]
    \centering
    \includegraphics[width=\linewidth]{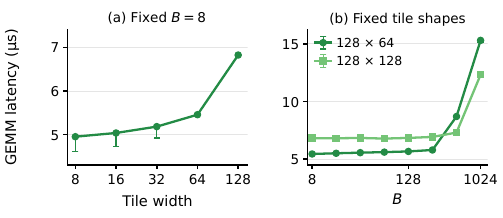}
    \caption{Pure NVFP4 GEMM latency. (a) Varying tile width at fixed $B=8$. (b) Varying $B$ with two fixed tile shapes.}
    \label{fig:mext_tile_headroom}
\end{figure}

\textbf{Residue compensation can use unused space within tiles and idle GPU resources to keep latency overhead low.} After the operand swap, result-tile columns correspond to tokens; we refer to the tile size along this dimension as tile width.
Figure~\ref{fig:mext_tile_headroom} examines how tile width and input size affect pure NVFP4 GEMM latency.
When the base and residue rows together fit within the tile width, compensation uses otherwise unused positions without increasing the number of tiles.
If the extension exceeds that capacity, a wider tile is another option.
Figure~\ref{fig:mext_tile_headroom}a fixes $B=8$ and increases tile width from 8 to 64; GEMM latency rises from 4.95 to 5.46\,$\mu$s, or 10.2\%.
Enlarging tiles to accommodate residues may therefore limit tile-count growth at a modest latency cost.

When the extension requires more tiles, the added tasks may still use parallel capacity left idle by the original workload.
Figure~\ref{fig:mext_tile_headroom}b fixes the input and output channel counts at $K=N=4096$ and varies $B$, giving a transposed result with 4096 rows and $B$ columns.
For output-channel tile size $T_N$ and tile width $T_B$, the logical tile count is
\begin{equation}
    \label{eq:mext_logical_tiles}
    n_{\mathrm{tiles}}=
    \left\lceil\frac{N}{T_N}\right\rceil
    \left\lceil\frac{B}{T_B}\right\rceil.
\end{equation}
Each tile covers 128 output channels and 64 token positions, dividing the output-channel dimension into 32 tiles.
Increasing $B$ from 128 to 256 raises the number of tiles along the token dimension from 2 to 4.
The total tile count thus grows from 64 to 128, both below the GPU's 148 SMs.
Measured latency rises by only 2.2\%, from 5.68 to 5.80\,$\mu$s.

As the input grows further, wider tiles can reduce the number of tasks to schedule.
At $B=512$, the preceding configuration produces 256 tiles.
Increasing tile width to 128 reduces this count to 128 and latency from 8.70 to 7.31\,$\mu$s.
This is consistent with executing the tiles in fewer waves.
Narrower tiles expose more independent tasks, whereas wider tiles fit more base and residue outputs and limit the increase in tile count.

\begin{figure}[!htbp]
    \centering
    \includegraphics[width=0.85\linewidth,trim=5.8bp 5.2bp 6bp 2.9bp,clip]{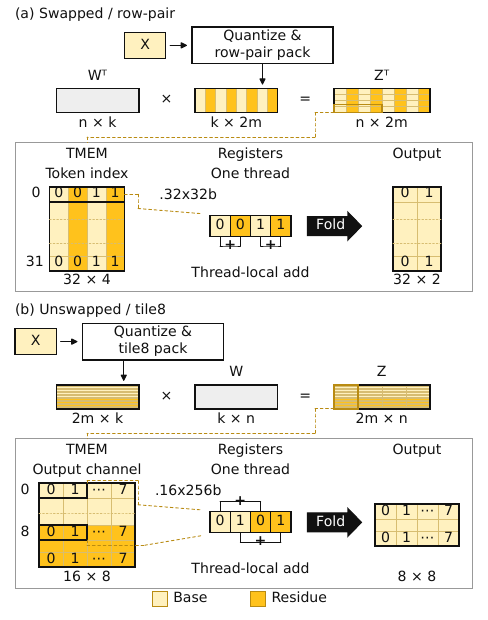}
    \caption{M-Ext packing and output folding.
    Swapped row-pair and unswapped tile8 layouts match TMEM loads to thread-local base--residue pairs.
    Insets show representative fragments.}
    \label{fig:mext_axis_swap}
\end{figure}

\textbf{Matching activation packing to TMEM loads keeps each base--residue pair in the same thread's registers.} Figure~\ref{fig:mext_axis_swap} shows the paired layouts.
On B200, Tensor Cores accumulate into Tensor Memory (TMEM), and the epilogue loads those accumulators into registers~\cite{nvidia_ptx_n.d.}.
The TMEM load shape determines which accumulator values each thread receives.
Each thread adds the pair in FP32 before the output cast.

In the swapped layout, the quantizer interleaves each base row with its residue row.
Their outputs occupy adjacent columns in the accumulator tile in TMEM.
A \texttt{.32x32b} TMEM load brings the corresponding values into thread-local registers for Fold.
For shapes better served by unswapped tiles, the quantizer packs eight base rows followed by their eight residue rows.
A \texttt{.16x256b} TMEM load places each pair in the same thread's registers.

After Fold, the epilogue writes the combined output to global memory, either through shared memory using a Tensor Memory Accelerator (TMA) store or directly from registers.
Figure~\ref{fig:mext_execution_ablation}b compares on-chip Fold against writing the expanded BF16 output and combining each pair with a separate Fold kernel.
With identical prepared operands, $N=K=4096$, and fixed $128\times64$ tiles, on-chip Fold gives $1.34$--$1.41\times$ speedups for $M=1$--$64$.

\textbf{M-Ext jointly autotunes tile shapes, activation packing, TMEM loads, and output stores.} For each combination of $M$, $N$, and $K$, it measures compatible configurations and selects the fastest, including activation packing, GEMM, and Fold in the timing.

\begingroup
\setlength{\parskip}{6pt plus 4pt minus 1pt}

\subsection{ResAttn Execution}
\label{sec:resattn_execution}
ResAttn applies M-Ext's paired computation to decode attention and combines the compensated scores before softmax.
Its execution must preserve the existing KV-cache layout and expose parallelism for long sequences.

\begingroup
\clubpenalties 3 10000 10000 0
\textbf{ResAttn pairs the query base and residue computations and folds their scores before softmax} (Figure~\ref{fig:resserve_overview}).
A separate query quantization kernel generates and packs FP4 base and residue rows from BF16 queries before the attention kernel runs.
The attention kernel directly uses the cached FP4 keys in the product of queries and keys (QK).
As in M-Ext, ResAttn swaps the operands so that query base and residue outputs occupy columns of the score tile.
It adds the corresponding scores on-chip in FP32 before applying online softmax~\cite{dao_flashattention_2022}.
\par
\endgroup

\textbf{ResAttn preserves the paged NVFP4 layout of both K and V.} Both operands retain scale groups along the head dimension.
This grouping matches the reduction dimension of QK, allowing direct FP4 execution.
In the product of probabilities and values (PV), however, the reduction dimension follows sequence positions rather than the head dimension.
Direct block-scaled FP4 PV would require grouping V scales along that dimension instead.
To retain the existing cache layout, ResAttn converts V from NVFP4 to FP8 on-chip.
ResAttn computes PV with FP8 operands, using FP32 accumulation and normalization.

\textbf{ResAttn adopts KV sequence splitting to expose more parallel work.} This is useful when the batch and KV-head dimensions provide too few thread blocks~\cite{dao_flashdecoding_2023,ye_flashinfer_2025a}.
Each split computes local attention outputs and softmax statistics; reduction combines these outputs using the statistics to preserve normalization across the full sequence.
More splits can improve parallelism, but also add storage for partial outputs, synchronization, and reduction work.
ResAttn jointly selects the split count and reduction method to balance these costs (Figure~\ref{fig:resserve_overview}).
Preset rules use batch size, KV-head count, sequence length, and the GPU's SM count to select the configuration without a timing search during serving.

\par
\endgroup

\subsection{Selective Prefill Execution}
\label{sec:prefill_compensation}
\label{sec:channel_selection}
\label{sec:large_m_kext}
\label{sec:stage_execution}
\textbf{Selective prefill controls added computation by compensating only a subset of channels.} Prefill processes more tokens concurrently, making the computational overhead of full compensation more pronounced.
Prior work shows that channels differ in quantization sensitivity and that activation statistics help identify channels worth protecting~\cite{lin_awq_2024,zhao_atom_2024}.
ResServe uses block-local selection for block-scaled FP4, with implementations for both NVFP4 and MXFP4.
Calibration uses mean absolute activation as a heuristic salience score and selects a fixed number of channels within each local group.
Inference reuses these selections.
Our current NVFP4 and MXFP4 implementations use 8-channel selection groups for partial compensation ratios of 0.125, 0.25, and 0.5.
Both also support full compensation at 1.0.
M-Ext is designed for full-residue compensation in decoding, while selective compensation controls the added work in prefill.

\textbf{Selected residue channels and matching weight rows can be incorporated into the original GEMM.} As shown in the upper-right panel of Figure~\ref{fig:resserve_overview}, we append quantized residues as activation columns and the corresponding quantized weights as weight rows:
\begin{equation}
    \label{eq:kext_augmented}
    \begin{aligned}
        \mathbf{Y} &\approx
        \begin{bmatrix}\hat{\mathbf{X}} & \hat{\mathbf{R}}_{\mathcal{S}}\end{bmatrix}
        \begin{bmatrix}\hat{\mathbf{W}} \\ \hat{\mathbf{W}}_{\mathcal{S}}\end{bmatrix} \\
        &= \hat{\mathbf{X}}\hat{\mathbf{W}}
        + \hat{\mathbf{R}}_{\mathcal{S}}\hat{\mathbf{W}}_{\mathcal{S}}.
    \end{aligned}
\end{equation}
This concatenation extends the GEMM reduction dimension, hence the name K-Ext.
A single FP4 GEMM directly sums the base and compensation terms of Eq.~\ref{eq:residue_compensation}.
Offline, we append the selected floating-point weight rows to the original weights, then quantize using the extended block layout.
Here, $\hat{\mathbf{W}}_{\mathcal{S}}$ denotes the selected weights quantized in this extended layout.
At runtime, the fused quantizer generates the base and selected residues in the corresponding concatenated layout.
For a selected channel fraction $\rho$, the effective reduction dimension is
\[
K_{\mathrm{eff}}=K(1+\rho).
\]

\begingroup
\clubpenalties 3 10000 10000 0
\textbf{Block-local selection with a fixed number of channels per group lets quantization directly produce the extended activation.} Global selection can leave threads holding different numbers of selected channels, requiring runtime reordering to form contiguous operands (Figure~\ref{fig:reordering_free}).
Even with offline selection indices, each new input still requires the corresponding indexing and data movement.
ResServe selects the same number from every group, making output offsets predictable.
The fused quantizer computes residues in registers from the original inputs and their FP4 base values.
It extracts selected residues and computes independent scales for contiguous residue blocks.
It then quantizes and packs them directly into the extension, avoiding a separate channel-reordering step.
Appendix Algorithm~\ref{alg:residue_quant} details the NVFP4 implementation.

\par
\endgroup
\begingroup
\setlength{\intextsep}{12pt}
\begin{figure}[!htbp]
    \centering
    \includegraphics[width=\linewidth,trim=18pt 10pt 18pt 10pt,clip]{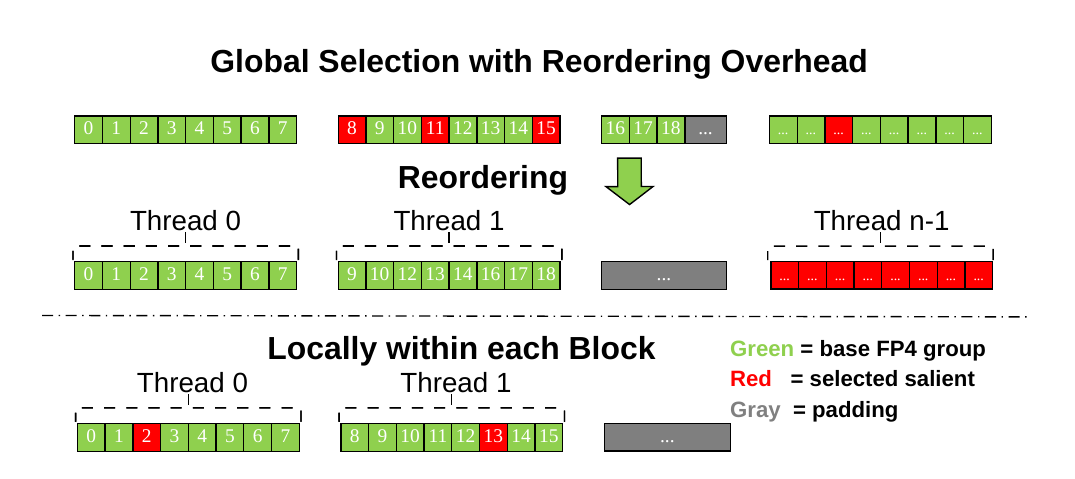}
    \caption{Channel selection under SIMT.
    Global selection uses runtime channel reordering (top); block-local selection with a fixed number of channels per group supports fused packing (bottom).}
    \label{fig:reordering_free}
\end{figure}
\endgroup

\subsection{System Integration}
\label{sec:implementation}
To support selective prefill compensation, we extend NVIDIA TensorRT Model Optimizer~\cite{nvidiacorporation_nvidia_2026} to save each layer's selected channels and execution mode during export.
The vLLM loader restores this information together with the corresponding weight layout.
Weight preparation is performed once during export, avoiding runtime weight reordering.
Prefill and decode share the base weight.
Partial compensation additionally stores the selected weight rows for the extended GEMM (Appendix~\ref{app:storage}).
An optional offline search assigns layer-wise ratios under a compensation budget (Appendix~\ref{app:ratio_planning}).

The online runtime is built on vLLM~\vllmver{}.
Both linear paths build on FlashInfer~\cite{ye_flashinfer_2025a}; M-Ext adds residue input layouts, tactic tuning for each input shape, and the output fold of Section~\ref{sec:mext_codesign}.
ResAttn reuses the existing KV cache layout, preserves the prefill attention path, and replaces only the decode attention computation.
The baseline uses TensorRT-LLM's decode attention kernel through FlashInfer~\cite{nvidia_nvidia_n.d.}.

\section{Evaluation}
\label{sec:eval}

\subsection{Experimental Setup}
\label{sec:experimental_setup}
We primarily evaluate on NVIDIA B200 (SM100), with additional RTX 5090 (SM120) NVFP4 linear-kernel results in Appendix~\ref{app:sm120}.
Task evaluations cover Qwen3-4B/8B/32B, Qwen3.8-27B, and Swift-1.5-Qwen3.8-27B (Swift-27B)~\cite{yang_qwen3_2025,qwenteam_qwen38max_2026,ukisai_swift_2026}.
We compare NVFP4 and MXFP4 M-Ext with same-format plain FP4 and W4A8 implementations from FlashInfer (FI) and TensorRT-LLM (TRT).
Plain denotes W4A4 execution without residue compensation.
Precision comparisons use the same original weights and BF16 activations.
Linear timing includes online activation quantization and GEMM/Fold.
Appendix~\ref{app:linear_protocol} specifies backend formats and tuning settings.

\subsection{Kernel Evaluation}
\label{sec:kernel_eval}

\subsubsection{Linear Kernels}
\label{sec:mext_w4a8}
\label{sec:operator_precision}
\label{sec:linear_latency}
\begin{figure}[!htbp]
    \centering
    \includegraphics[width=\linewidth]{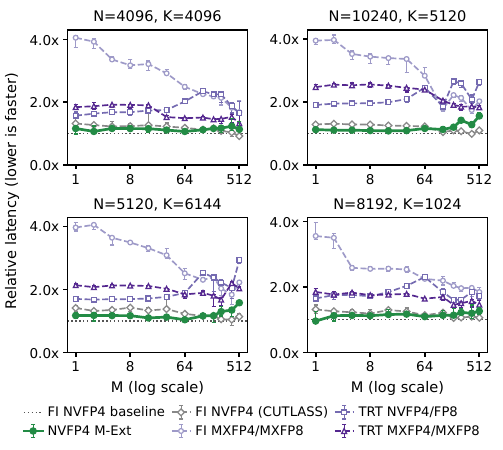}
    \caption{B200 linear latency relative to FI NVFP4 CuTeDSL (lower is better).
    \textbf{Green curves show our NVFP4 M-Ext.} FI NVFP4 CUTLASS is shown for reference.}
    \label{fig:mext_b200_latency}
\end{figure}

For $M$ from 1 to 64, NVFP4 M-Ext achieves median speedups of $1.53$--$1.83\times$ over TRT NVFP4/FP8.
Over the same row counts, MXFP4 M-Ext achieves median speedups of $1.55$--$1.96\times$ over TRT MXFP4/MXFP8.
NVFP4 M-Ext incurs 3.1--7.8\% median overhead over plain NVFP4.
The NVFP4 summaries cover 22 projection shapes; Figure~\ref{fig:mext_b200_latency} shows four representative shapes.
Appendix Figures~\ref{fig:nvfp4_b200_full} and~\ref{fig:mxfp4_b200_latency} provide the per-shape sweeps.

At full compensation, M-Ext lowers execution cost relative to K-Ext (Appendix Figure~\ref{fig:full_compensation}).
With identical NVFP4 base and residue values, M-Ext is $1.47$--$1.48\times$ faster than full K-Ext for $M=1$--$128$ and $N=K=4096$.
This comparison includes online activation quantization and uses independently selected kernel configurations.

M-Ext also reduces output error on Qwen3-32B against an FP32 reference (Appendix Table~\ref{tab:linear_precision}).
The comparison uses 72 paired real inputs and includes both weight and activation quantization error.
Pooled normalized MSE (NMSE) divides the summed squared error by the summed squared reference values across all evaluated inputs.
NVFP4 M-Ext lowers NMSE by 16.6\% relative to plain NVFP4 and 24.8\% relative to TRT NVFP4/FP8.
MXFP4 M-Ext reduces NMSE by 20.3\% relative to plain MXFP4, approaching the output precision of MXFP4/MXFP8.

\begin{figure}[!tb]
    \centering
    \includegraphics[width=\linewidth]{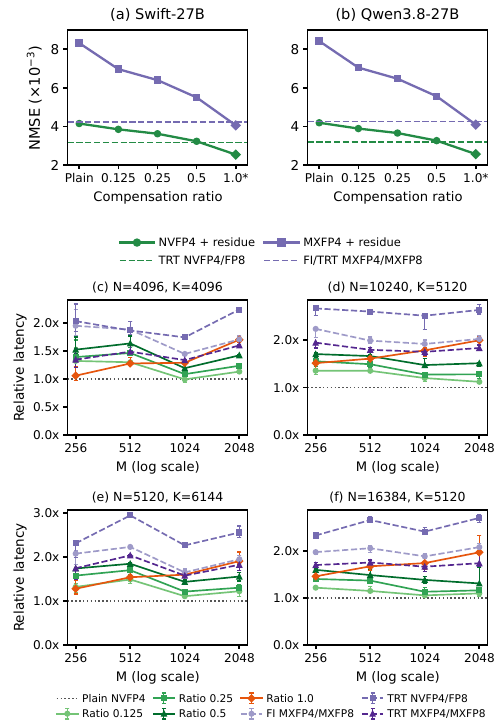}
    \caption{Selective compensation: (a--b) linear NMSE; (c--f) B200 linear latency relative to FI plain NVFP4 (lower is better). (*) marks full compensation; ratio positions in (a--b) are categorical.}
    \label{fig:quality-ratio}
    \label{fig:kext_b200_latency}
\end{figure}

\label{sec:kext_latency}
\label{sec:eval_kext_quality}
Increasing the prefill compensation ratio reduces NMSE in both FP4 formats on Swift-27B and Qwen3.8-27B (Figure~\ref{fig:quality-ratio}(a--b)), using 192 matched prefill input groups per model.
At ratio 0.5, NVFP4 NMSE is less than 1.8\% above TRT NVFP4/FP8 on both models.
Full compensation reduces error below same-format W4A8 on both models and formats.

Selective compensation reduces prefill overhead relative to full compensation (Figure~\ref{fig:kext_b200_latency}(c--f)).
At $M=1024$ and $2048$, median latency across 17 shapes is $1.11$--$1.14\times$ plain NVFP4 at ratio 0.125, versus $1.69$--$1.97\times$ for full compensation.
Appendix Figure~\ref{fig:kext_b200_full} provides the per-shape sweeps.

\subsubsection{Decode Attention}
\label{sec:resattn_eval}
\begin{figure}[!tb]
    \centering
    \includegraphics[width=\linewidth]{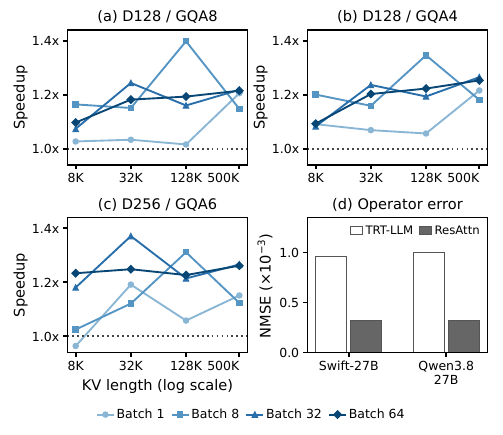}
    \caption{ResAttn versus native BF16-output TRT-LLM with page32 NVFP4 KV: (a--c) B200 speedup; (d) real-input pooled NMSE against the same dequantized KV reference.}
    \label{fig:resattn_latency}
\end{figure}

ResAttn achieves up to $1.40\times$ speedup over TensorRT-LLM, with a median speedup of $1.19\times$ across 48 configurations.
It is faster in 47 configurations (Figure~\ref{fig:resattn_latency}(a--c)).
Both kernels use identical BF16 queries and prepared page32 NVFP4 KV, and directly produce BF16 output.
The TRT baseline converts BF16 queries to FP8 before attention.
Timing covers query conversion and attention, excluding cache construction and KV append.

ResAttn also lowers pooled NMSE by 66.3\% on Swift-27B and 68.1\% on Qwen3.8-27B across 192 paired real-input attention instances per model (Figure~\ref{fig:resattn_latency}(d)).
The FP32 attention reference uses the same NVFP4 KV, isolating attention computation error from KV quantization loss.
Using FP32 attention on the original BF16 Q, K, and V as the reference, ResAttn reduces total NMSE relative to TRT-LLM by 14.98\% on Swift-27B and 16.24\% on Qwen3.8-27B.

\subsection{End-to-End Latency}
\label{sec:e2e_latency}
\begin{table}[t]
    \centering\small
    \setlength{\tabcolsep}{3pt}
    \renewcommand{\arraystretch}{0.94}
    \caption{HTTP latency (ms) with NVFP4 KV, averaged across three independent server processes.
    ResServe prefill ratios are 0 (NVFP4) and 0.25 (MXFP4).
    Appendix~\ref{app:e2e_protocol} and Appendix Table~\ref{tab:e2e-nvfp4-ratios} give the protocol and NVFP4 ratio sweep.}
    \label{tab:e2e-latency}
    \vskip 0.10in
    \begin{tabular*}{\linewidth}{@{\extracolsep{\fill}}llrrr@{}}
        \toprule
        Format & Method & TTFT & TPOT & E2E latency \\
        \midrule
        \multicolumn{5}{l}{\emph{Qwen3-8B}} \\
        NVFP4 & W4A4 & 90.90 & 2.940 & 840.48 \\
        & W4A8 & 142.87 & 3.830 & 1119.39 \\
        & ResServe & 91.58 & 3.048 & 868.76 \\
        \addlinespace[2pt]
        MXFP4 & W4A4 & 88.59 & 2.747 & 789.04 \\
        & W4A8 & 117.05 & 4.646 & 1301.81 \\
        & ResServe & 96.83 & 2.932 & 844.49 \\
        \midrule
        \multicolumn{5}{l}{\emph{Qwen3-32B}} \\
        NVFP4 & W4A4 & 274.27 & 8.520 & 2446.96 \\
        & W4A8 & 686.33 & 17.863 & 5241.45 \\
        & ResServe & 273.96 & 8.521 & 2446.91 \\
        \addlinespace[2pt]
        MXFP4 & W4A4 & 260.06 & 7.948 & 2286.71 \\
        & W4A8 & 450.20 & 17.599 & 4938.03 \\
        & ResServe & 309.27 & 8.601 & 2502.50 \\
        \bottomrule
    \end{tabular*}
\end{table}

We evaluate whether kernel efficiency benefits carry over to full-model serving with M-Ext and ResAttn.
The low-load HTTP evaluation covers NVFP4 and MXFP4 on Qwen3-8B and Qwen3-32B, with one SM100 GPU per model.
Requests contain 8,192 input and 256 output tokens, with Poisson arrivals at 1 request/s.
Table~\ref{tab:e2e-latency} reports time to first token (TTFT), mean time per output token after the first (TPOT), and complete request latency.
The W4A8 baselines use TensorRT-LLM's NVFP4/FP8 and MXFP4/MXFP8 GEMMs for NVFP4 and MXFP4 weights, respectively.

With selective prefill compensation disabled, NVFP4 ResServe adds 3.4\% request-latency overhead on Qwen3-8B and retains nearly unchanged latency on Qwen3-32B relative to W4A4.
MXFP4 ResServe uses the quality evaluation's prefill compensation ratio of 0.25, adding 7.0\% and 9.4\% overhead over same-format W4A4 on Qwen3-8B and Qwen3-32B, respectively.
Relative to the evaluated W4A8 baselines, ResServe reduces request latency by 22.4--53.3\% with NVFP4 and 35.1--49.3\% with MXFP4.

\begin{table}[t]
    \centering
    \small
    \setlength{\tabcolsep}{3pt}
    \caption{System accuracy (\%): LCB pass@1; HMMT/AIME avg@16.
    BF16 references use BF16 KV; FP4 rows use NVFP4 KV.
    ResServe uses prefill ratios 0 (NVFP4) and 0.25 (MXFP4).
    Bold marks the better same-format score.}
    \label{tab:quality-system}
    \vskip 0.15in
    \begin{tabular*}{\linewidth}{@{\extracolsep{\fill}}llrrr@{}}
        \toprule
        \multicolumn{2}{l}{Configuration} & LCB & HMMT & AIME \\
        \midrule
        \emph{Qwen3-4B} & BF16 & 36.57 & 51.25 & 63.13 \\
        NVFP4 & W4A4 & 23.71 & 31.88 & 50.00 \\
        & ResServe & \textbf{28.86} & \textbf{37.92} & \textbf{58.75} \\
        MXFP4 & W4A4 & 10.57 & 14.58 & 28.13 \\
        & ResServe & \textbf{20.86} & \textbf{33.96} & \textbf{50.83} \\
        \midrule
        \emph{Qwen3-8B} & BF16 & 45.14 & 55.83 & 68.33 \\
        NVFP4 & W4A4 & 24.29 & 43.13 & 56.25 \\
        & ResServe & \textbf{35.14} & \textbf{48.96} & \textbf{61.04} \\
        MXFP4 & W4A4 & 12.00 & 22.08 & 36.88 \\
        & ResServe & \textbf{27.14} & \textbf{40.63} & \textbf{55.21} \\
        \midrule
        \emph{Qwen3-32B} & BF16 & 51.14 & 64.38 & 75.63 \\
        NVFP4 & W4A4 & 42.29 & 55.21 & \textbf{71.46} \\
        & ResServe & \textbf{46.29} & \textbf{61.67} & 71.25 \\
        MXFP4 & W4A4 & 31.43 & 45.83 & 57.71 \\
        & ResServe & \textbf{38.00} & \textbf{52.50} & \textbf{67.08} \\
        \midrule
        \emph{Qwen3.8-27B} & BF16 & 84.00 & 79.38 & 84.17 \\
        NVFP4 & W4A4 & \textbf{79.43} & \textbf{80.83} & \textbf{88.75} \\
        & ResServe & 77.71 & 78.96 & 88.33 \\
        MXFP4 & W4A4 & 70.86 & 66.25 & 72.92 \\
        & ResServe & \textbf{74.86} & \textbf{72.29} & \textbf{78.75} \\
        \midrule
        \emph{Swift-27B} & BF16 & 83.43 & 76.25 & 82.08 \\
        NVFP4 & W4A4 & 76.57 & 75.42 & 81.46 \\
        & ResServe & \textbf{77.71} & \textbf{78.33} & \textbf{85.42} \\
        MXFP4 & W4A4 & 65.71 & 62.50 & 67.71 \\
        & ResServe & \textbf{77.43} & \textbf{69.17} & \textbf{75.83} \\
        \bottomrule
    \end{tabular*}
\end{table}

\begin{figure}[!b]
    \centering
    \includegraphics[width=\linewidth]{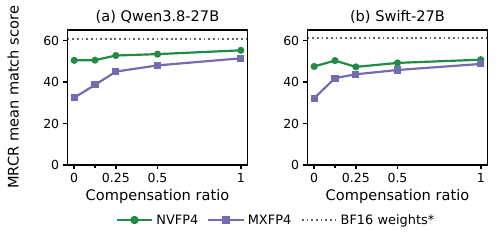}
    \caption{Mean MRCR matching scores (0--100) on Qwen3.8-27B and Swift-27B.
    All configurations use NVFP4 KV and TRT-LLM attention.
    Dotted lines (*) are BF16-weight references with the same KV.}
    \label{fig:mrcr_ratio_kv}
\end{figure}

\subsection{Task Ablation and Quality}
\label{sec:system_eval}
\label{sec:mext_task_quality}
We first isolate M-Ext on the Hard350 subset of LiveCodeBench (LCB) v6~\cite{jain_livecodebench_2024}, using BF16 KV and the same attention implementation.
Within each FP4 format, plain and M-Ext share quantized weights, prompts, and generation settings.
Selective prefill is disabled in this comparison.
Averaged across five models, relative score gains over same-format plain inference are 7.70\% for NVFP4 and 18.18\% for MXFP4.
Total output tokens, including reasoning and final answers across all problems, decrease in all ten model--format pairs, with mean reductions of 4.54\% for NVFP4 and 11.81\% for MXFP4.

\label{sec:e2e_eval}
With NVFP4 KV, adding M-Ext to ResAttn improves scores in all ten model--format pairs, with mean relative gains of 8.12\% for NVFP4 and 30.97\% for MXFP4 across five models.
Appendix Table~\ref{tab:lcb-full-sweep} reports component ablations and output-token savings.

System evaluations use a fixed prefill compensation ratio of 0 for NVFP4 and 0.25 for MXFP4 across all five models.
The MXFP4 setting is motivated by the retrieval gains from selective linear compensation in Figure~\ref{fig:mrcr_ratio_kv}, discussed later in this section.
Table~\ref{tab:quality-system} reports LCB, HMMT November 2025, and AIME 2026 results under these configurations.
With NVFP4 KV in both configurations, ResServe enables M-Ext and ResAttn, while same-format plain inference uses TensorRT-LLM attention.
For each model and task, compared methods use the same prompts, thinking mode, sampling parameters, and output budget.
LCB reports pass@1 from one generation per problem.
Mathematics evaluations use MathArena~\cite{dekoninck2026matharena} and report avg@16, the mean correctness over 16 samples per problem.

Across the five models, ResServe improves LCB accuracy over same-format plain inference with mean relative gains of 15.04\% for NVFP4 and 53.57\% for MXFP4.
On LCB, total output tokens decrease by an average of 3.18\% for NVFP4 and 9.16\% for MXFP4 relative to the same baselines.
Mean relative mathematics gains range from 6.02\% to 9.15\% for NVFP4 and 33.34\% to 50.23\% for MXFP4.
MXFP4 improves mathematics scores in all ten model--task pairs.
To assess sensitivity to which questions are included, we perform 10,000 paired bootstrap resamples of questions.
For both FP4 formats on all three tasks, the 95\% confidence intervals for mean score differences in percentage points across the five models lie above zero.
This supports ResServe's average quality gains on code generation and mathematics.

ResServe achieves higher scores than the evaluated W4A8 baselines on Qwen3-4B and Qwen3-8B across all three tasks and both FP4 formats.
Scores are broadly comparable on the 27B and 32B models (Appendix Table~\ref{tab:quality-system-w4a8}).

\label{sec:eval_accuracy}
Selective linear compensation also improves retrieval on a fixed 300-item MRCR~\cite{openai_openai_2025a} 8-needle subset spanning 4--32K contexts.
With NVFP4 KV and TRT-LLM attention held fixed, ratio 0.5 improves scores over same-format plain inference across all five models and both FP4 formats.
Figure~\ref{fig:mrcr_ratio_kv} shows the two 27B models; Appendix Figure~\ref{fig:mrcr_five_models} provides all five.

\section{Discussion}
\label{sec:discussion}

\subsection{The Execution Cost of Precision Recovery}
M-Ext recovers precision at near-plain FP4 latency for the evaluated linear operators with few rows.
Both NVFP4 and MXFP4 M-Ext achieve lower latency than the evaluated W4A8 implementations.
NVFP4 M-Ext also reduces output error, and MXFP4 M-Ext approaches W4A8 output precision.
These results establish efficiently organized FP4 compensation as an alternative to increasing activation bit width for precision recovery.
With identical quantized values, M-Ext executes full compensation faster than K-Ext, showing that computation organization changes the cost of recovering the same residue information.
For larger prefill operations, selective compensation improves precision over plain FP4 and reduces execution cost relative to full compensation.

Low compensation cost carries through to full-model inference (Table~\ref{tab:e2e-latency}).
In the evaluated HTTP workload, the quality evaluation's configurations for both FP4 formats add less than 10\% request-latency overhead over same-format W4A4 and reduce latency relative to the evaluated W4A8 baselines.
Together with the independently measured task-quality gains, these results demonstrate the practical value of FP4 residual compensation.

\subsection{Task Quality and Output Length}
Precision recovery can improve task quality and reduce output length.
Prior work finds that some aggressive quantization settings reduce reasoning accuracy and increase generation length~\cite{lotfi_quantized_reasoning_2026}.
In the BF16-KV LiveCodeBench ablation, linear compensation improves mean task scores while reducing total output tokens across all model--format pairs.
The full system also improves mean LCB accuracy while reducing output tokens on average in both FP4 formats.
These results make generation length relevant alongside per-token execution cost when evaluating quantized reasoning.

The benefits of compensation depend on the inference stage and task.
Selective compensation improves long-context retrieval, while M-Ext provides low-cost full-residue compensation during reasoning generation.
The gains from additional prefill compensation in code generation vary across models, motivating task-aware selection of the compensation configuration.

\section{Related Work}

Quantization methods reduce weight and activation error~\cite{frantar_optq_2023,lin_awq_2024,xiao_smoothquant_2023,zhao_atom_2024}.
Higher-precision activation paths use wider floating-point formats~\cite{micikevicius_fp8_2022,mishra_recipes_2025}, W4A8 execution~\cite{lin_qserve_2025}, or selective precision allocation~\cite{saxena_resq_2025b,liu_micromix_2026a}.
Other work varies precision between prefill and decode~\cite{lu_mixquant_2026} in disaggregated serving~\cite{zhong_distserve_2024}.
Quantized residues have been used to correct weights~\cite{wang_fairy2i_2025} and activations~\cite{meng_arcquant_2026}.
ResServe couples compensation with operator-specific execution, using on-chip result combination and block-local selection to control its runtime cost.
Low-precision attention reduces computation cost~\cite{zhang_sageattention2_2025,zhang_sageattention3_2025a}, while KV-cache compression reduces storage~\cite{liu_kivi_2024a,hooper_kvquant_2024}.
ResAttn applies M-Ext's compensation execution principle to attention, combining compensation scores before softmax while reusing the existing paged NVFP4 cache.

FlashInfer and FlashDecoding++ adapt execution to input shapes~\cite{ye_flashinfer_2025a,hong_flashdecoding_2024}, while Flash-Decoding exploits parallelism across the KV sequence for attention~\cite{dao_flashdecoding_2023}.
ResServe selects kernel configurations for each input shape, accounting for compensation layouts, Fold costs, and attention reduction costs.

\section{Conclusion}
ResServe improves block-scaled FP4 task quality with low compensation cost.
The system adapts residue computation to the matrix dimensions, combines results on-chip, and selectively compensates prefill channels.
It integrates NVFP4/MXFP4 linear computation with paged NVFP4 KV caching.
Full-model evaluations demonstrate higher mean coding and mathematics accuracy, alongside retrieval gains from selective linear compensation.
On code generation, these quality gains are accompanied by average reductions in output tokens.
Low-load NVFP4 serving measurements establish low integration overhead, retaining near-baseline request latency with selective prefill disabled.
Efficiently organizing additional low-precision computation offers a practical way to retain low-bit execution efficiency while improving LLM performance on downstream tasks.
We plan to extend this principle to grouped GEMM in mixture-of-experts (MoE) models, exploring low-cost compensation under different expert loads.

\bibliographystyle{mlsys2025}
\begingroup
\setlength{\bibsep}{1pt plus 1pt}
\bibliography{references}
\endgroup

\clearpage
\pdfbookmark[0]{Appendices}{appendices}
\appendix
\twocolumn[\section*{Appendices}]
\makeatletter
\setlength{\@fptop}{0pt}
\setlength{\@dblfptop}{0pt}
\makeatother

\section{NVFP4 Linear Kernels on RTX 5090}
\label{app:sm120}

We extend the main text's evaluation of NVFP4 linear kernels to RTX 5090 (SM120), measuring M-Ext and selective prefill overhead and comparing with ARCQuant.
We measure four Qwen3-8B projections against FlashInfer's auto-selected plain NVFP4 baseline.
Timing includes online activation quantization and GEMM/Fold; Section~\ref{app:sm120_protocol} gives the measurement protocol.

\subsection{Linear-Layer Latency}
\label{app:legacy_linear}
\begin{figure}[!t]
    \centering
    \includegraphics[width=\linewidth]{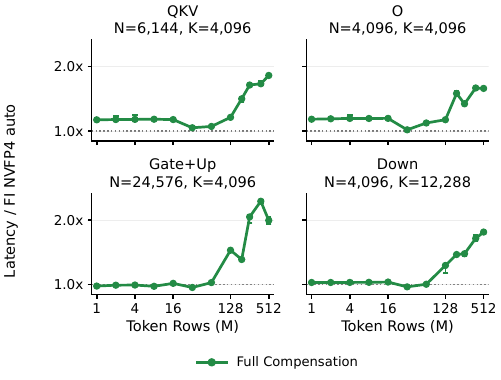}
    \caption{Full-compensation latency relative to plain FlashInfer NVFP4 auto on RTX 5090; lower is better.}
    \label{fig:mext_latency}
\end{figure}

M-Ext remains close to plain NVFP4 latency at small row counts (Figure~\ref{fig:mext_latency}).
For $M=1$--$64$, its latency is $0.95$--$1.20\times$ that of plain FlashInfer, with a cross-configuration median of $1.04\times$.

\subsection{Selective Prefill and ARCQuant Comparison}
\label{app:baselines}
\label{app:baseline_latency}
\begin{figure}[!t]
    \centering
    \includegraphics[width=\linewidth]{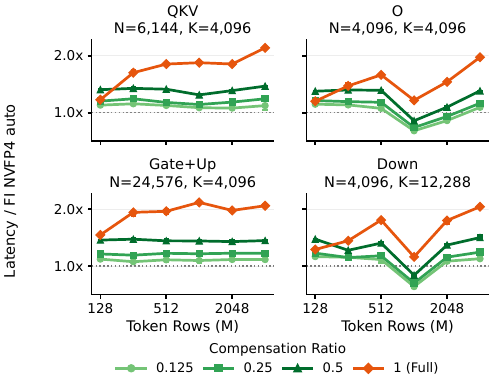}
    \caption{Selective prefill latency relative to plain FlashInfer NVFP4 auto on RTX 5090; lower is better.
    Full compensation is included for reference.}
    \label{fig:linear_latency_baselines}
\end{figure}

Selective prefill reduces compensation overhead at larger row counts (Figure~\ref{fig:linear_latency_baselines}).
Median latencies relative to plain FlashInfer are $1.12\times$, $1.19\times$, and $1.40\times$ for compensation ratios 0.125, 0.25, and 0.5, respectively, versus $1.81\times$ for full compensation.

\begin{figure*}[t]
    \centering
    \includegraphics[width=\textwidth]{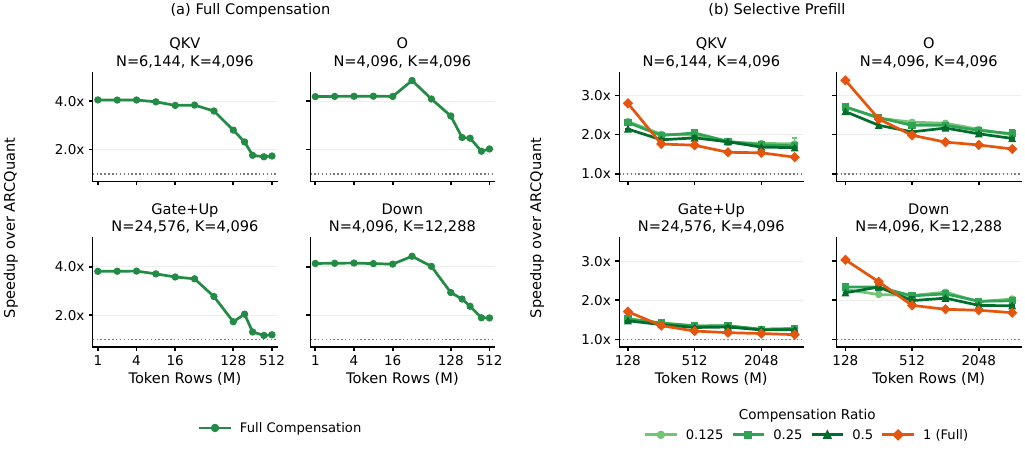}
    \caption{ResServe speedup over ARCQuant on RTX 5090 at matching compensation ratios; higher is better.}
    \label{fig:arcquant_comparison}
\end{figure*}

\textbf{ResServe is faster than ARCQuant at matching compensation ratios}~\cite{meng_arcquant_2026} (Figure~\ref{fig:arcquant_comparison}).
M-Ext achieves a median speedup of $4.05\times$ for $M=1$--$64$.
K-Ext is faster in all 72 selective configurations, with speedups of $1.24$--$2.71\times$ (median $1.99\times$).

Both paths include online activation quantization and GEMM.
ARCQuant's implementation computes a global activation scale for each quantization call.
Our timing excludes the global amax reduction and scale computation but includes activation prescaling and channel reordering.
This exclusion favors ARCQuant by omitting its per-call scale preparation overhead.
FlashInfer and ResServe use autotuning; ARCQuant uses its fixed compiled kernels.

\section{Implementation Details}
\label{app:implementation_details}

\subsection{Block-Local Selection and Residue Packing}
\label{app:channel_selection}
Partial compensation selects a fixed number of channels within each 8-channel group, using ratios $\rho\in\{0.125,0.25,0.5\}$.
Each layer uses one ratio, with every group contributing $k=8\rho$ channels.
Full compensation includes all channels.
Calibration scores channels by their mean absolute activation $m_i$ and selects the top-$k$ within each group $G_g=\{8g,\dots,8g+7\}$:
\begin{equation}
    \mathcal{S} = \bigcup_g \mathrm{TopK}\bigl(\{m_i : i \in G_g\},\, k\bigr).
\end{equation}
Calibration uses 1,024 samples with a sequence-length cap of 2,048, batch size 8, and seed 42.
One 8-bit mask per group records the selected positions and remains fixed at inference.

\begin{figure}[t]
    \centering
    \includegraphics[width=\linewidth]{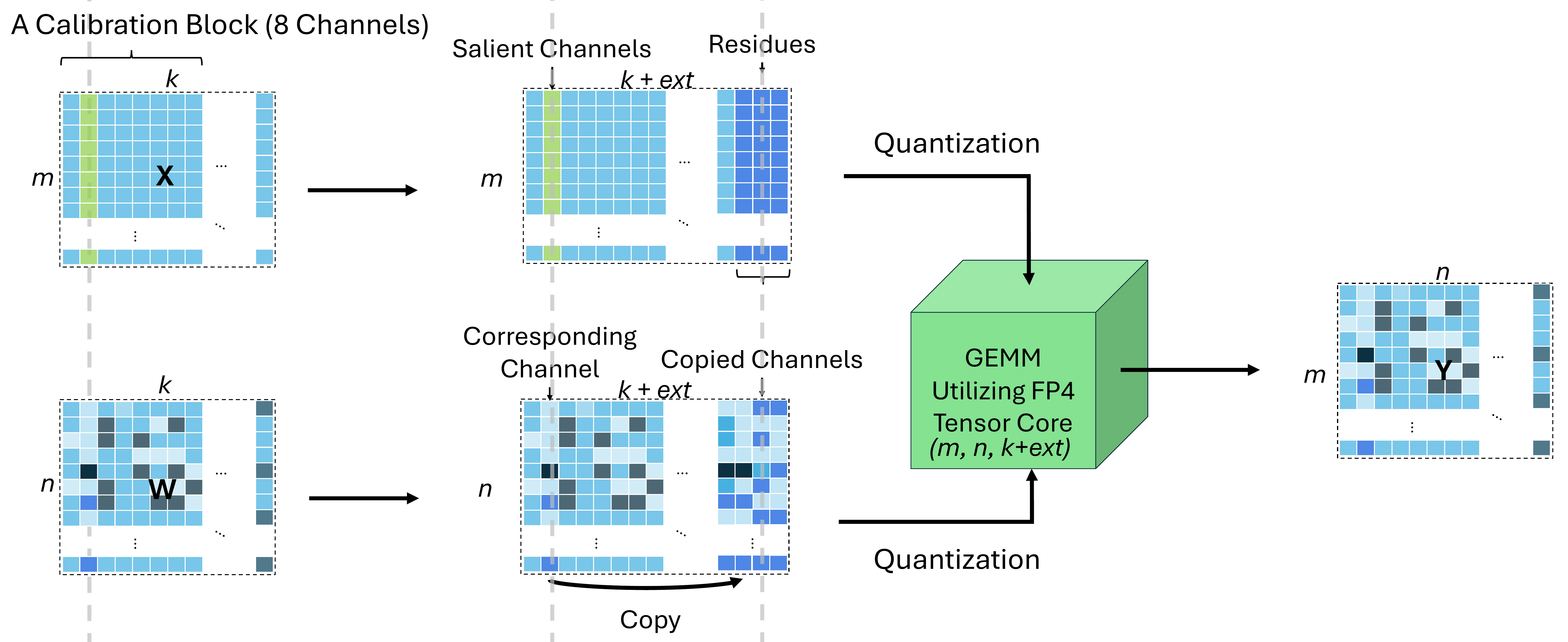}
    \caption{
        Selective K-Ext pairs runtime residue channels with weight rows prepared offline in an augmented FP4 GEMM.
    }
    \label{fig:residuequant_overview}
\end{figure}

The selected operands are quantized as $\hat{\mathbf{R}}_{\mathcal{S}}=\mathrm{QDQ}(\mathbf{R}_{\mathcal{S}})$ and $\hat{\mathbf{W}}_{\mathcal{S}}=\mathrm{QDQ}(\mathbf{W}_{\mathcal{S}})$.
The same $\mathcal{S}$ pairs the selected residue channels with their corresponding weight rows.
Figure~\ref{fig:residuequant_overview} illustrates the augmented GEMM of Eq.~\ref{eq:kext_augmented} for NVFP4 prefill with partial compensation.

\subsection{NVFP4 Quantize-and-Extend}
\label{app:fused_quantization}
\label{app:nvfp4_scales}
\textbf{NVFP4 combines an FP8 block scale with an FP32 tensor scale.} Each 16-element block shares an E4M3 scale; the tensor scale supplies the common range~\cite{nvidia_pretraining_2026,cook_four_2026}.
The maximum magnitudes of E2M1 and E4M3 are 6 and 448, respectively.
When the tensor scale is determined from the current input range, each block $B_i$ in tensor $\mathbf{X}$ uses
\begin{equation}
    \mathrm{Max}_i = \max_{x \in B_i} |x|,
    \qquad
    s_i = \frac{\mathrm{Max}_i}{6},
    \qquad
    \alpha = \frac{\max_i s_i}{448},
\end{equation}
where $\alpha$ is the tensor-level scale.
The stored FP8 block scale encodes the normalized value $s_i/\alpha$.
Its effective dequantization scale includes $\alpha$:
\begin{equation}
    s_{i,\mathrm{FP8}} = \mathrm{round}_{\mathrm{E4M3}}(s_i/\alpha),
    \qquad \tilde{s}_i = \alpha s_{i,\mathrm{FP8}}.
\end{equation}
For $x\in B_i$, FP4 quantization and reconstruction use this effective scale:
\begin{equation}
    x_{\mathrm{FP4}} = \mathrm{round}_{\mathrm{E2M1}}(x/\tilde{s}_i),
    \qquad \hat{x} = \tilde{s}_i x_{\mathrm{FP4}}.
\end{equation}

\textbf{The fused kernel generates base and residue operands in one pass.} It receives a precomputed tensor scale and computes block scales at runtime.
The NVFP4 vLLM integration loads calibrated tensor scales from the checkpoint and reuses them during inference.
We describe the pack8 implementation below, which assigns eight input elements to each thread.
Two adjacent threads use a warp shuffle to compute the maximum magnitude for their shared 16-element scale block.
The kernel quantizes inputs to FP4, then decodes and rescales them to compute residues in registers without writing reconstructed activations to memory.

The selected residues form a separate NVFP4 operand with its own block scales and share the base operand's tensor scale.
Each thread contributes $k$ residues, so a scale group spans $16/k$ consecutive threads.
Warp shuffles gather scaled residues into eight-value chunks, which are then quantized and packed into 32-bit words for coalesced writes.
Algorithm~\ref{alg:residue_quant} summarizes this partial-compensation mapping; \textsc{ReduceMax} reduces over the indicated number of adjacent lanes within one warp.

\begin{algorithm}[t]
    \caption{Residue NVFP4 quantization: eight inputs per thread.}
    \label{alg:residue_quant}
    \footnotesize
    \begin{algorithmic}[1]
        \Require \Type{bf16}~$x[8]$ (8 activations; or fp16), \Type{uint8}~$\mathit{mask}$, \Type{fp32}~$\alpha$ (tensor scale)
        \Ensure (to global memory) main: $8$ \Type{fp4} $+$ \Type{fp8} SF (1 per $2$ lanes); residue: $8\rho$ \Type{fp4} $+$ \Type{fp8} SF (1 per $16/k$ lanes)
        \AlgComment{on-lane bf16/fp16 $\to$ fp32}
        \State \Type{fp32}~$\mathit{xf}[8] = \mathrm{cast}(x)$
        \AlgComment{per-16 amax: per-lane max, then reduce over $2$ lanes ($8$ elts each)}
        \State \Type{fp32}~$\mathit{amax} = \textsc{ReduceMax}_{2}(\max_i |\mathit{xf}[i]|)$
        \AlgComment{\textsc{BlockScale}: fp8 $\mathit{sf}$ (one per $2$ lanes) and the fp32 multiplier $\mathit{scale}$}
        \State \Type{fp8}~$\mathit{sf}$, \Type{fp32}~$\mathit{scale} = \textsc{BlockScale}(\alpha, \mathit{amax})$
        \State \Kw{if} $\mathit{lane} \bmod 2 == 0$:\quad store $\mathit{sf}$
        \AlgComment{\textsc{QDQ}: fp4 codes $+$ fp32 dequant}
        \State \label{ln:qdq}\Type{fp4}~$\mathit{q}[8]$, \Type{fp32}~$\mathit{x\_q}[8] = \textsc{QDQ}(\mathit{scale} \cdot \mathit{xf})$; store $\mathit{q}$
        \AlgComment{residue over all 8 channels, in registers}
        \State \label{ln:residue}\Kw{for} $i = 0..7$:\quad \Type{fp32}~$\mathit{r}[i] = \mathit{xf}[i] - \mathit{x\_q}[i] / \mathit{scale}$
        \State \label{ln:select}\Type{int}~$k = \textsc{popcount}(\mathit{mask})$;\quad \Type{fp32}~$\mathit{r\_sel}[k] = \textsc{Select}(\mathit{r}, \mathit{mask})$
        \AlgComment{per-16 residue amax: reduce over $16/k$ lanes ($k$ residues each)}
        \State \label{ln:rscale}\Type{fp32}~$\mathit{r\_amax} = \textsc{ReduceMax}_{16/k}(\max_i |\mathit{r\_sel}[i]|)$
        \AlgComment{\textsc{BlockScale}: fp8 $\mathit{sf\_r}$ (one per $16/k$ lanes) and fp32 $\mathit{scale\_r}$}
        \State \Type{fp8}~$\mathit{sf\_r}$, \Type{fp32}~$\mathit{scale\_r} = \textsc{BlockScale}(\alpha, \mathit{r\_amax})$
        \State \label{ln:rscale-end}\Kw{if} $\mathit{lane} \bmod (16/k) == 0$:\quad store $\mathit{sf\_r}$
        \AlgComment{scale residues before gathering across lanes}
        \State \Type{fp32}~$\mathit{r\_scaled}[k] = \mathit{scale\_r} \cdot \mathit{r\_sel}$
        \AlgComment{gather eight consecutive residues for each writer lane}
        \State \Type{fp32}~$\mathit{chunk}[8] = \textsc{WarpGather8}(\mathit{r\_scaled})$
        \AlgComment{quantize eight values to FP4 and pack into one 32-bit word}
        \State \label{ln:rq}\Type{uint32}~$\mathit{c} = \textsc{QuantizePack8}(\mathit{chunk})$
        \State $w = 32\rho$;\quad \Kw{if} $\mathit{lane} < w$:\quad store $\mathit{c}$
    \end{algorithmic}
\end{algorithm}

\section{Evaluation Protocols}
\label{app:evaluation_details}

\subsection{Kernel Timing}
\label{app:linear_protocol}

\paragraph{Linear layers}
Linear-layer timing includes online activation quantization and GEMM/Fold to BF16 output.
On B200, M-Ext and K-Ext cover 22 and 17 projection shapes, respectively, using shared synthetic BF16 inputs and weights.
In selective prefill comparisons, full compensation denotes a compensation ratio of 1.0.
Timing uses CUDA Graphs with cold-buffer rotation.

M-Ext and plain FI CuTeDSL use the same FP4 format; W4A8 baselines use FI/TRT MXFP4/MXFP8 or TRT NVFP4/FP8.
The NVFP4 and TRT FP8 linear operator tests precompute global activation scales from each test input's maximum absolute value.
Timing excludes this computation but includes online block-scale generation for NVFP4 and MXFP8.

Kernel selection differs across the evaluated backends.
M-Ext and plain CuTeDSL use FI autotuning, while K-Ext uses FI automatic GEMM selection.
FI MXFP4/MXFP8 uses its default 1-SM tactic.
TRT NVFP4/FP8 uses native selection, whereas TRT MXFP4/MXFP8 uses FI tactic tuning.
FI CUTLASS NVFP4 is an additional backend reference.

The B200 projection-shape sweeps normalize latency to one common baseline per format.
The NVFP4 sweep uses FI NVFP4 CuTeDSL; the MXFP4 sweep uses FI MXFP4 stock.
Points show the median paired latency ratio across five independent processes; bars span the minimum and maximum.
Cross-shape speedup summaries take the median of paired baseline/M-Ext ratios per shape, then the median across 22 distinct shapes.

\label{app:sm120_protocol}
The RTX 5090 linear comparisons use cold-L2 CUDA Graph timing across three independent processes.
Global-scale preparation and offline weight packing are excluded for all methods.
Each point is the median paired latency ratio across processes; bars show their range.
Reported summary medians aggregate these per-configuration ratios.

\paragraph{Decode attention}
\label{app:resattn_protocol}
Decode-attention timing includes query conversion and attention, but excludes cache construction and KV append.
ResAttn and the TRT-LLM baseline with native BF16 output receive identical BF16 queries and prepared page32 NVFP4 KV with global scales of 1.0.
Timing uses warm buffers; the baseline uses FlashInfer's native dispatcher and default TRT-LLM selection.
KV-head counts are 8, 8, and 4 in panel order.

Each speedup point is the median of three within-process baseline/ResAttn latency ratios.
Overall summaries take the median across configurations using all process records.

\begin{figure}[!t]
    \centering
    \includegraphics[width=\linewidth,viewport=0 0 237.6 144,clip]{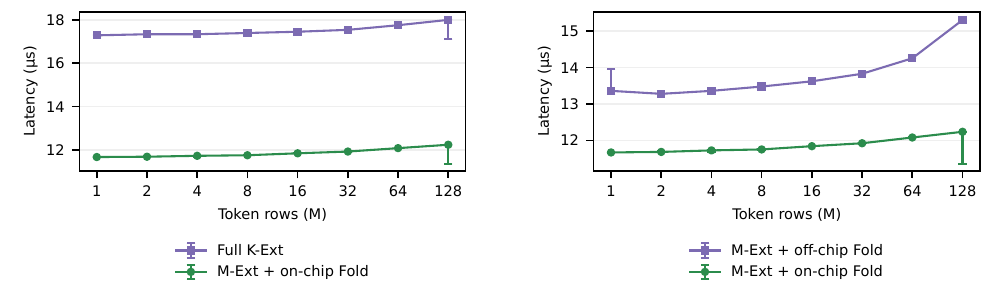}
    \caption{Full NVFP4 compensation on SM100 with $N=K=4096$: M-Ext versus K-Ext with offline duplicated weights.
    Timing includes online activation quantization, packing, and GEMM/Fold.
    Points show medians across three processes; bars show their range.}
    \label{fig:full_compensation}
\end{figure}

\paragraph{Full-compensation comparison}
Figure~\ref{fig:full_compensation} compares complete NVFP4 paths on SM100 with $N=K=4096$, using synthetic BF16 inputs and weights.
Both paths use pack8 quantization and identical base, residue, and weight codes and scales.
K-Ext concatenates along $K$ and duplicates weights offline; M-Ext uses paired rows.
Each path independently selects from $128\times64$ and $128\times128$ tiles with three pipeline stages and single-CTA clusters.
K-Ext also selects whether to swap operands.

Timing includes online quantization, packing, and GEMM/Fold, excluding global-scale preparation and offline weight processing.
Cold-buffer CUDA Graph measurements use three independent processes, with per-process medians paired before aggregating speedups.
Outputs are checked against a common FP32 reference reconstructed from the represented FP4 values.

\subsection{Operator Precision}
\label{app:linear_precision_protocol}
\label{app:swift_prefill_precision}

Linear precision measures error against FP32 matrix multiplication of the original BF16 operands, with TF32 disabled.
Within each cohort, all implementations receive identical captured activations and original BF16 weights, prepared in each backend's supported quantized representation.
Plain FP4 and same-format M-Ext share quantized weights; W4A8 uses its backend's representation.
Output error includes both weight and activation quantization.

The linear studies cover QKV, output, gate/up, and down projections using real decode and prefill inputs.
Decode uses 72 inputs from Qwen3-32B; prefill uses 192 input groups per model from Swift-27B and Qwen3.8-27B.
The prefill inputs come from AIME, HMMT, and MRCR prompts.
The prefill study excludes hybrid linear-attention projections.

Attention precision uses two references, with and without KV quantization loss.
The main reference is FP32 attention on the original query and the same dequantized NVFP4 KV, measuring computation error across the complete path.
The additional BF16-KV reference also includes cache quantization loss.
Both comparisons use 192 matched decode instances per model from Swift-27B and Qwen3.8-27B.

Precision metrics aggregate squared errors before normalization.
Pooled MSE divides total squared error by the number of output elements; pooled NMSE divides it by total reference energy.
Both attention references use pooled NMSE.

\subsection{Reasoning Tasks}
\label{app:lcb_protocol}
\label{app:system_quality}

The reasoning evaluations compare linear compensation under BF16 KV and system components under NVFP4 KV.
The latter uses page size 32 and global K/V scales of 1.0.
TRT-LLM uses native BF16 output, and ResAttn replaces decode attention.
M-Ext applies at scheduled row counts up to 128, including eligible small prefill chunks.

LCB reports pass@1, while mathematics scores report avg@16: mean correctness over 16 samples per problem.
LCB v6 uses the 350 hard problems in release\_v6, with official scoring and one generation per problem.
\href{https://huggingface.co/datasets/MathArena/hmmt_nov_2025}{HMMT November 2025} and \href{https://huggingface.co/datasets/MathArena/aime_2026}{AIME 2026} each contain 30 problems from MathArena~\cite{dekoninck2026matharena}, scored with its official parser.
Length-limited outputs remain in the scores.
LCB token savings include reasoning and final answers across all 350 problems.

For system evaluations, relative score gains and LCB output-token reductions are computed against each model's same-format plain baseline, then averaged equally across the five models.
For the BF16-KV M-Ext ablation, relative score gains and output-token reductions are computed against each model's same-format plain baseline, then averaged equally across the five models.
For the NVFP4-KV component ablation, mean relative score gains from adding M-Ext use ResAttn alone as the baseline, with selective prefill disabled in both configurations.

Generation settings are matched across methods within each model.
Qwen3-4B/8B/32B use thinking mode at temperature 0.6; Qwen3.8-27B and Swift-27B use medium-effort thinking at temperature 1.0.
Both LCB and mathematics runs use top-p 0.95 and top-k 20.
LCB uses seed 42 for all five models.
The 16 mathematics samples use seeds 42--57 for Qwen3 and 0--15 for the two 27B models, with the same seed set across compared methods.
Their context limits are 40,960 tokens for Qwen3 and 262,144 for the two 27B models.

\textbf{We additionally evaluate W4A8 baselines on code generation and mathematics tasks.} Table~\ref{tab:quality-system-w4a8} compares these baselines with plain W4A4 and ResServe across five models, using the task cohorts, scoring definitions, model-specific sampling settings, and seed sets described above.

\subsection{Retrieval Task}
\label{app:mrcr_protocol}

MRCR~\cite{openai_openai_2025a} compares linear compensation while holding the KV format and attention implementation fixed.
All arms, including the BF16-weight reference, use page32 NVFP4 KV with global K/V scales of 1.0 and TRT-LLM native BF16-output attention.
Ratio 0 denotes uncompensated inference; nonzero ratios apply the checkpoint's compensation configuration in both prefill and decode, without separately enabling decode M-Ext.
Ratio 1 retains the checkpoint's full-compensation path.
ResAttn is not used.

All configurations use the same 300-item, 8-needle subset spanning 4--32K contexts.
We use greedy, non-thinking generation with a 2,048-token output limit and one run per configuration.
Scores use the official prefix check and SequenceMatcher, averaged equally over all 300 answers and scaled to 0--100.
Length-stopped and empty answers remain included.

\subsection{End-to-End Latency}
\label{app:e2e_protocol}

We measure client-side streaming HTTP latency with \texttt{vllm bench serve}, using one SM100 GPU per model.
Synthetic requests contain exactly 8,192 input and 256 output tokens, with EOS ignored.
Arrivals follow a Poisson process at 1 request/s with seed 17 and a client concurrency cap of 32.
Each of three independent server processes serves 32 warmup requests followed by 128 measured requests.
We report the mean of the three process means.

We evaluate NVFP4 and MXFP4 weights with page32 NVFP4 KV caches in a controlled comparison.
The baseline uses upstream vLLM linear auto dispatch with a TRT BF16-output attention adapter.
ResServe at ratio 0 uses the same checkpoint with M-Ext and ResAttn; nonzero ratios use selective residue exports.
M-Ext follows the production row-count threshold rather than a strict prefill/decode switch.
FlashInfer autotuning is enabled for the W4A4 and ResServe configurations.

The runtime uses vLLM 0.26.0, Torch 2.11.0, adapted FlashInfer 0.7.0, and CUTLASS DSL 4.8.0, with full and piecewise CUDA Graphs.
Chunked prefill is enabled and prefix caching disabled.
We integrate TensorRT-LLM's NVFP4/FP8 GEMM, corresponding to the TRT NVFP4/FP8 baseline in the kernel evaluation, into vLLM for the W4A8 system comparison.
The NVFP4 W4A8 measurements use TensorRT-LLM 1.3.0rc24 with static per-tensor FP8 activation scales and GEMM configurations selected by TensorRT-LLM's native heuristic.
The MXFP4 W4A8 baseline uses TensorRT-LLM's MXFP4/MXFP8 GEMM with online block-scaled MXFP8 activation quantization and FlashInfer tactic tuning.
W4A8 baselines are measured separately using the same framework versions and HTTP workload protocol.

HTTP latency includes queuing, execution, and client/transport overhead.
Output throughput is limited by the offered request rate, not maximum serving capacity.

\section{Additional Evaluation Results}
\label{app:additional_results}

\subsection{Kernel Results}
\label{app:mxfp4_latency}
\label{app:full_linear_results}

Table~\ref{tab:linear_precision} reports linear-operator error on Qwen3-32B under the protocol in Appendix~\ref{app:linear_precision_protocol}.

\begin{table}[t]
    \centering
    \small
    \setlength{\tabcolsep}{3pt}
    \caption{Linear-operator error on Qwen3-32B (lower is better). Bold values indicate the lowest error within each format.}
    \label{tab:linear_precision}
    \vskip 0.15in
    \begin{tabular*}{\linewidth}{@{\extracolsep{\fill}}llrr@{}}
        \toprule
        Implementation & Compute & MSE & NMSE \\
        \midrule
        \multicolumn{4}{@{}l}{\emph{NVFP4}} \\
        FI plain & W4A4 & 0.295745 & 0.008401 \\
        \textbf{M-Ext} & W4A4 & \textbf{0.246697} & \textbf{0.007007} \\
        TRT NVFP4/FP8 & W4A8 & 0.328123 & 0.009320 \\
        \midrule
        \multicolumn{4}{@{}l}{\emph{MXFP4}} \\
        FI plain & W4A4 & 0.790377 & 0.022450 \\
        \textbf{M-Ext} & W4A4 & 0.630280 & 0.017903 \\
        FI MXFP4/MXFP8 & W4A8 & \textbf{0.625482} & \textbf{0.017767} \\
        TRT MXFP4/MXFP8 & W4A8 & \textbf{0.625482} & \textbf{0.017767} \\
        \bottomrule
    \end{tabular*}
\end{table}

\begin{table*}[t]
    \centering\small
    \setlength{\tabcolsep}{2pt}
    \caption{LCB Hard350 pass@1 scores and output-token savings (\%), one generation per problem.
    Bold marks the best score within each model, FP4 format, and KV block.}
    \label{tab:lcb-full-sweep}
    \label{tab:lcb-token-savings}
    \vskip 0.15in
    \begin{tabular*}{\textwidth}{@{\extracolsep{\fill}}llcrrrrrrrrrr@{}}
        \toprule
        & & & \multicolumn{2}{c}{Qwen3-4B} & \multicolumn{2}{c}{Qwen3-8B} & \multicolumn{2}{c}{Qwen3-32B} & \multicolumn{2}{c}{Qwen3.8-27B} & \multicolumn{2}{c}{Swift-27B} \\
        \cmidrule(lr){4-5}\cmidrule(lr){6-7}\cmidrule(lr){8-9}\cmidrule(lr){10-11}\cmidrule(l){12-13}
        Weights & Configuration & \shortstack{Prefill\\ratio} & Score & \shortstack{Tokens\\saved} & Score & \shortstack{Tokens\\saved} & Score & \shortstack{Tokens\\saved} & Score & \shortstack{Tokens\\saved} & Score & \shortstack{Tokens\\saved} \\
        \midrule
        BF16 & BF16 KV & --- & 36.57 & --- & 45.14 & --- & 51.14 & --- & 84.00 & --- & 83.43 & --- \\
        \midrule
        \multicolumn{13}{l}{\textit{(a) BF16 KV: M-Ext ablation}} \\
        \addlinespace[2pt]
        NVFP4 & Baseline & 0 & 25.14 & 0.00 & 32.00 & 0.00 & 41.71 & 0.00 & \textbf{77.43} & 0.00 & 81.14 & 0.00 \\
        & Baseline + M-Ext & 0 & \textbf{28.86} & 6.04 & \textbf{36.29} & 4.52 & \textbf{45.43} & 5.25 & \textbf{77.43} & 3.23 & \textbf{82.29} & 3.67 \\
        \addlinespace[3pt]
        MXFP4 & Baseline & 0 & 19.71 & 0.00 & 17.43 & 0.00 & \textbf{35.14} & 0.00 & 73.14 & 0.00 & 72.57 & 0.00 \\
        & Baseline + M-Ext & 0 & \textbf{26.57} & 12.75 & \textbf{25.71} & 10.43 & 34.57 & 4.94 & \textbf{76.86} & 18.61 & \textbf{76.29} & 12.35 \\
        \midrule
        \multicolumn{13}{l}{\textit{(b) NVFP4 KV: component ablation}} \\
        \addlinespace[2pt]
        BF16 & TRT-LLM & --- & 32.86 & --- & 40.29 & --- & 46.86 & --- & 85.43 & --- & 84.57 & --- \\
        & ResAttn & --- & 33.14 & --- & 40.00 & --- & 48.29 & --- & 84.29 & --- & 85.14 & --- \\
        \addlinespace[3pt]
        NVFP4 & Baseline (TRT-LLM) & 0 & 23.71 & 0.00 & 24.29 & 0.00 & 42.29 & 0.00 & \textbf{79.43} & 0.00 & 76.57 & 0.00 \\
        & ResAttn & 0 & 26.00 & 8.13 & 30.29 & 1.06 & 41.71 & -3.33 & 76.86 & -0.67 & 76.57 & -5.87 \\
        & ResAttn + M-Ext & 0 & \textbf{28.86} & 11.31 & \textbf{35.14} & 6.25 & \textbf{46.29} & -1.25 & 77.71 & 0.68 & \textbf{77.71} & -1.12 \\
        \addlinespace[3pt]
        MXFP4 & Baseline (TRT-LLM) & 0 & 10.57 & 0.00 & 12.00 & 0.00 & 31.43 & 0.00 & 70.86 & 0.00 & 65.71 & 0.00 \\
        & ResAttn & 0 & 14.86 & 3.30 & 13.14 & 3.69 & 33.71 & -1.65 & 74.00 & 0.49 & 63.43 & -4.15 \\
        & ResAttn + M-Ext & 0 & \textbf{23.43} & 15.40 & 22.29 & 9.36 & 34.86 & 5.46 & \textbf{78.57} & 16.52 & 74.86 & 13.60 \\
        & ResAttn + M-Ext & 0.25 & 20.86 & 13.98 & \textbf{27.14} & 5.51 & \textbf{38.00} & 5.42 & 74.86 & 20.88 & \textbf{77.43} & 0.03 \\
        \bottomrule
    \end{tabular*}
\end{table*}

\begin{figure*}[p]
    \centering
    \includegraphics[width=\textwidth]{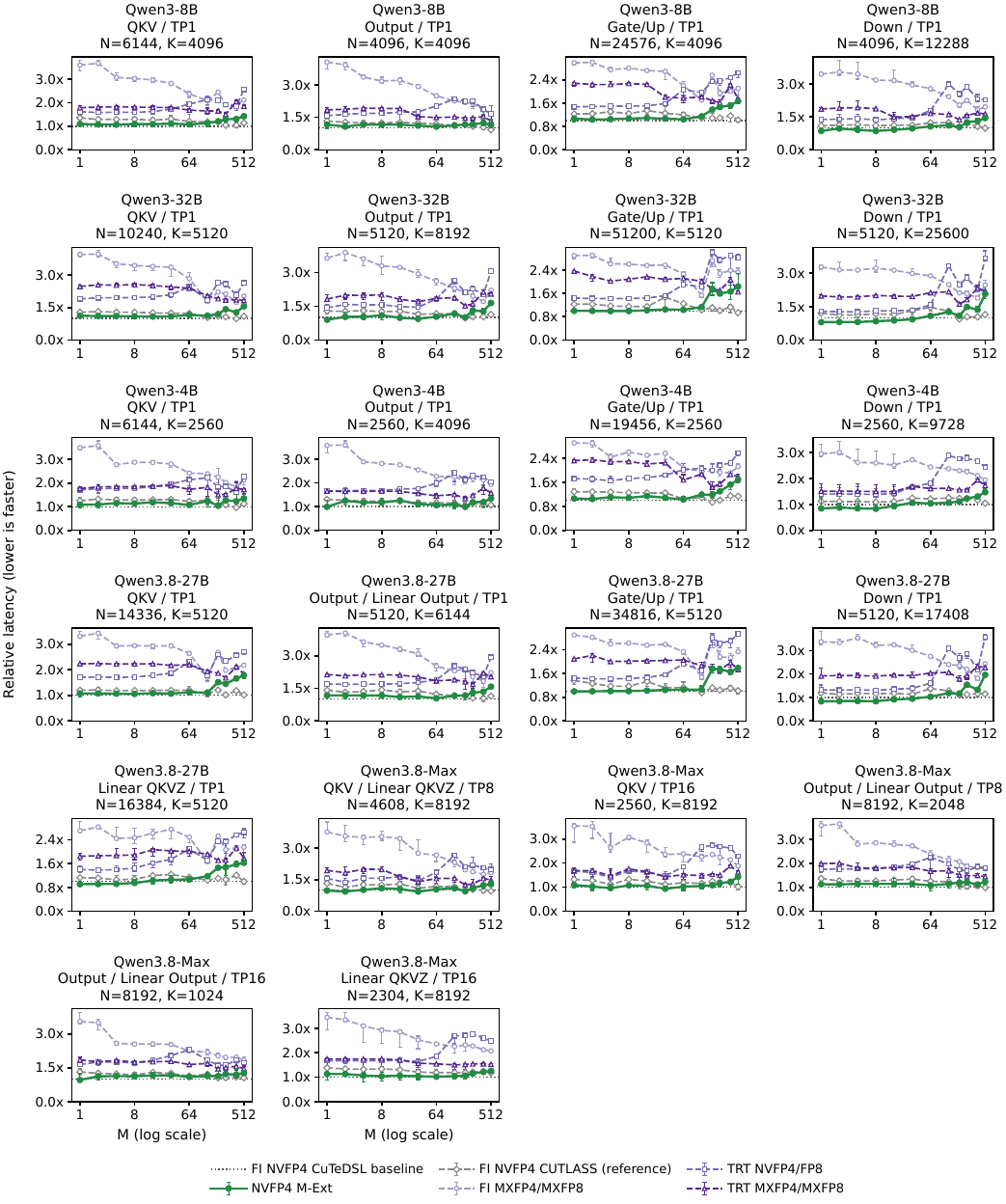}
    \caption{B200 NVFP4 M-Ext latency across 22 shapes, relative to plain FI NVFP4 CuTeDSL (lower is better).}
    \label{fig:nvfp4_b200_full}
\end{figure*}

\begin{figure*}[p]
    \centering
    \includegraphics[width=\textwidth]{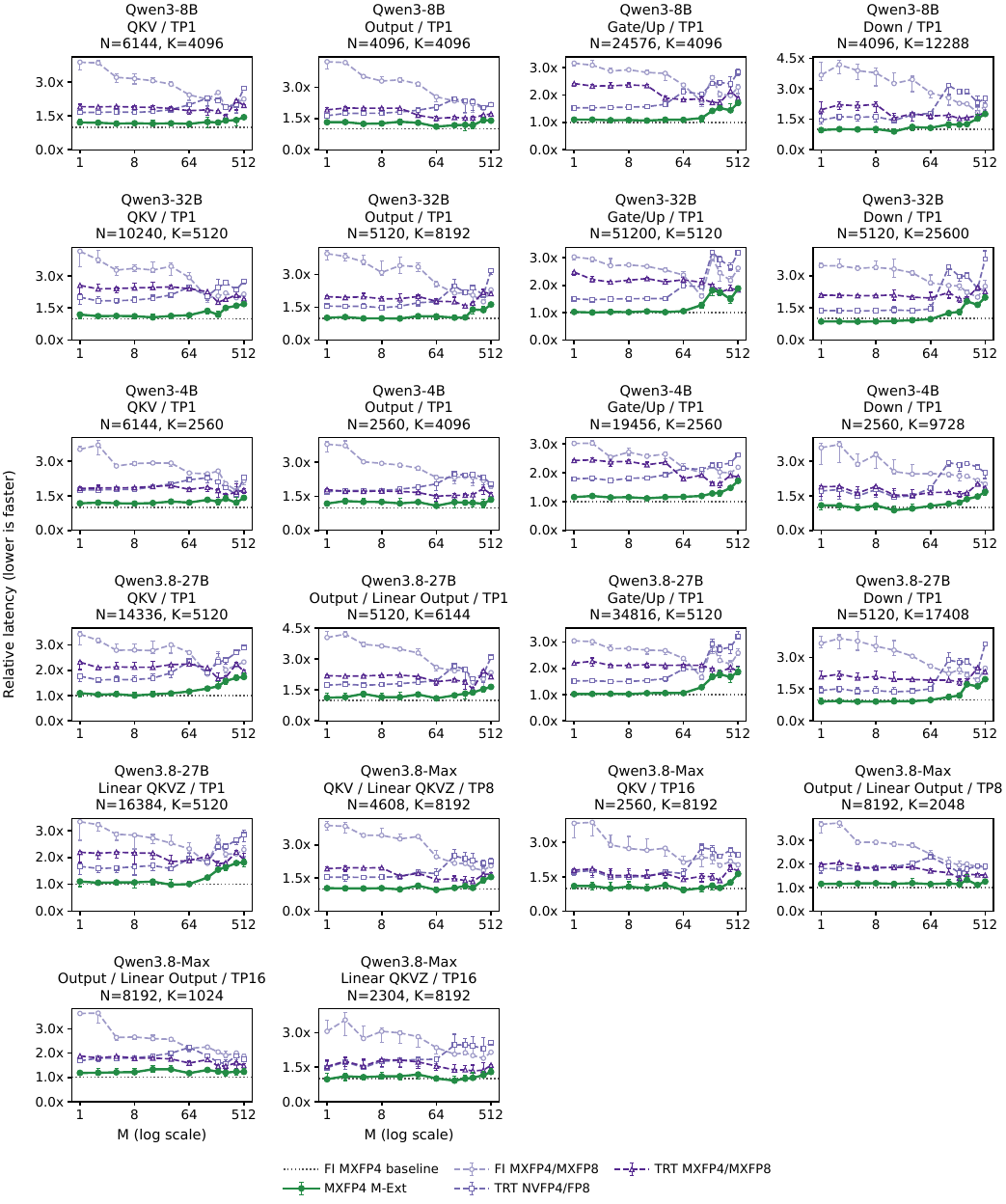}
    \caption{B200 MXFP4 M-Ext latency across 22 shapes, relative to plain FI MXFP4 (lower is better).}
    \label{fig:mxfp4_b200_latency}
\end{figure*}

\begin{figure*}[p]
    \centering
    \includegraphics[width=\textwidth]{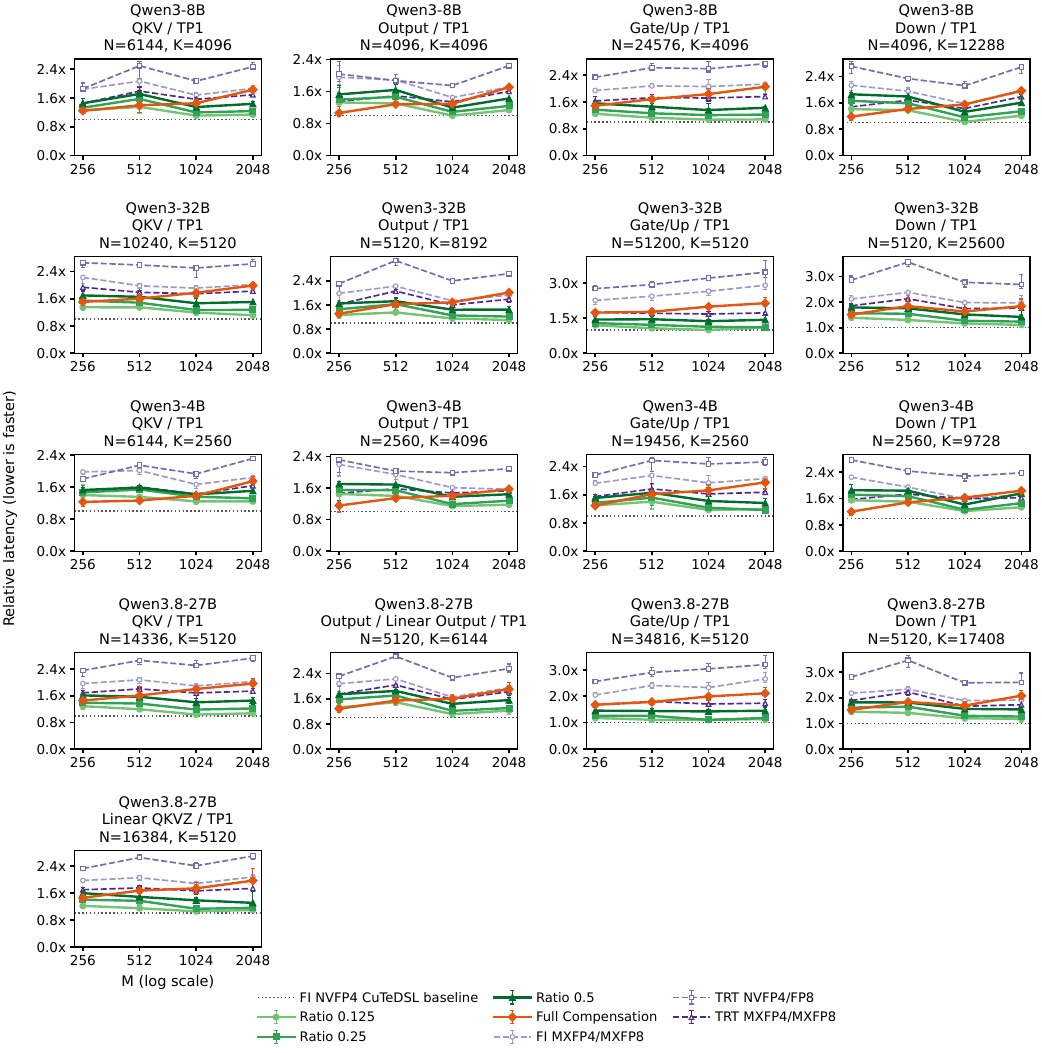}
    \caption{B200 selective prefill latency across 17 shapes, relative to plain FI NVFP4 CuTeDSL (lower is better).}
    \label{fig:kext_b200_full}
\end{figure*}

Figures~\ref{fig:nvfp4_b200_full}--\ref{fig:kext_b200_full} provide additional per-shape linear-latency results on B200.
Cross-shape summaries use the full measured shape sets described in Appendix~\ref{app:linear_protocol}.

Across $M=1$--$64$, the median MXFP4 M-Ext/plain latency ratio ranges from 1.06 to 1.14.
Over the same row counts, median speedups are $2.31$--$3.22\times$ over FI MXFP4/MXFP8 and $1.55$--$1.96\times$ over TRT MXFP4/MXFP8.

ResAttn also reduces total attention error relative to TRT-LLM when the reference uses the original BF16 KV, including cache quantization loss.
ResAttn reduces pooled NMSE from 0.004595 to 0.003907 on Swift-27B (14.98\%).
On Qwen3.8-27B, it reduces NMSE from 0.004706 to 0.003941 (16.24\%).

\subsection{LCB Ablation}
\label{app:lcb_results}

Table~\ref{tab:lcb-full-sweep} reports LCB pass@1 scores and output-token savings across five models under the protocol in Appendix~\ref{app:lcb_protocol}.
The BF16-KV block isolates M-Ext; the NVFP4-KV block compares M-Ext and ResAttn, with an additional MXFP4 row using selective prefill at ratio 0.25.

The MXFP4 prefill comparison holds M-Ext and ResAttn fixed.
Increasing the ratio from 0 to 0.25 improves three models' scores and lowers two; all five final system scores exceed the same-format baseline.
The uniform ratio follows the retrieval-based choice described in Section~\ref{sec:system_eval}.

Token savings include reasoning and final answers, relative to the same-model, same-format baseline within each KV block.
Negative values indicate longer outputs; dashes denote unreported savings.

\subsection{System Quality with W4A8 Baselines}
\label{app:w4a8_quality}

Table~\ref{tab:quality-system-w4a8} extends the main text's system comparison with NVFP4/FP8 and MXFP4/MXFP8 W4A8 baselines across all five models.
It retains the BF16 references, plain W4A4 results, and fixed ResServe configurations.
LCB reports Hard350 pass@1; HMMT November 2025 and AIME 2026 report average correctness over 16 samples per problem.
Appendix~\ref{app:lcb_protocol} specifies the additional baselines and their evaluation settings.

\subsection{Retrieval Quality Across Five Models}
\label{app:mrcr_nvfp4}

Figure~\ref{fig:mrcr_five_models} extends the main text's MRCR comparison to all five models using the protocol in Appendix~\ref{app:mrcr_protocol}.
Full compensation improves matching scores over same-format plain inference in all ten model--format pairs.
Gains range from 3.42 to 19.00 points for MXFP4 and from 1.23 to 4.82 points for NVFP4.

\begin{figure}[!htbp]
    \centering
    \includegraphics[width=\linewidth,trim=2bp 5bp 2bp 2bp,clip]{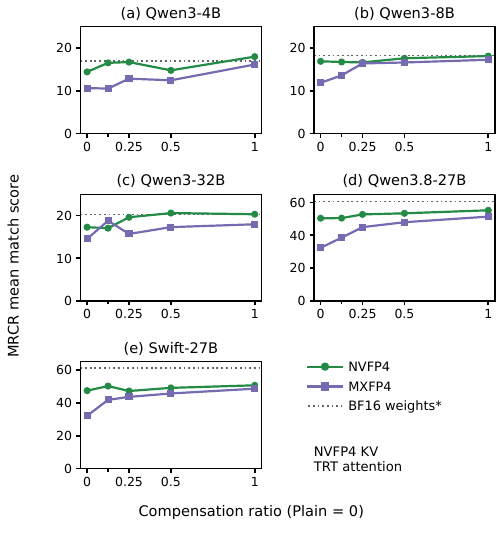}
    \caption{Mean MRCR matching scores (0--100) across five models.
    All configurations use NVFP4 KV and TRT-LLM attention.
    Dotted lines (*) are BF16-weight references with the same KV.
    Plain is at ratio 0; y-axis ranges differ.}
    \label{fig:mrcr_five_models}
\end{figure}

\subsection{Checkpoint Storage}
\label{app:storage}
Selective compensation trades additional checkpoint storage for precision recovery.
Table~\ref{tab:ratio_storage} reports complete checkpoint sizes for five models, relative to plain checkpoints of the same model and FP4 format.
Sizes count the logical bytes of deployment files, including weight shards, quantization metadata, channel masks, configuration, and tokenizer files.
This measures checkpoint storage rather than runtime GPU memory.
At ratio 0.125, storage increases by 7.67--10.66\% for NVFP4 and 7.51--10.57\% for MXFP4.
At ratio 0.25, the increases are 15.32--21.30\% and 14.99--21.12\%, respectively.
The main system evaluations use the plain checkpoint at NVFP4 ratio 0 and the expanded checkpoint at MXFP4 ratio 0.25.

Checkpoint growth differs from the channel selection ratio because unquantized tensors and fixed files remain unchanged, while alignment can add padding.
For Qwen3-4B MXFP4 at ratio 0.125, an extended width of 2,880 is padded to 2,944.

\section{Layer-Wise Compensation: Configuration and Evaluation}
\label{app:budgeted_compensation}

\subsection{Compensation-Ratio Planner}
\label{app:ratio_planning}
\label{sec:ratio_planning}
An optional offline planner selects each layer's NVFP4 prefill compensation ratio under an effective-bit budget.
Candidate ratios are 0, 0.125, 0.25, 0.5, and 1.0.
Calls with few rows use full-residue M-Ext independently of this plan.

Candidate scores estimate quantization damage using either gradient-weighted error in layer outputs or KL divergence between the output distributions of the reference and candidate models.
Gradient-weighted reconstruction also appears in BRECQ~\cite{li_brecq_2021}.
The KL option evaluates one layer or coupled layer group at a time, leaving other searchable layers unquantized.
It requires forward passes but no loss gradients.

The planner selects one ratio per layer to minimize the total score within the budget.
Effective-bit costs account for base and compensated FP4 computation, weighted by layer size; they do not represent physical storage.
As in budgeted mixed-precision allocation~\cite{yao_hawqv3_2021}, an offline integer linear program uses precomputed scores and costs, without additional forward passes.

\subsection{Reasoning Quality with AWQ}
\label{app:composition}
Residue compensation acts on activations and can be combined with offline weight optimization such as AWQ~\cite{lin_awq_2024}.

Table~\ref{tab:awq_reasoning} reports results combining AWQ with residue compensation for Qwen3-4B and Qwen3-8B at budgets 4.5, 5.0, and 5.5, with layer-wise ratios selected by the planner.
Both models run in thinking mode through vLLM with a BF16 KV cache.
The benchmarks are the full LiveCodeBench v6 set (LCB, pass@1)~\cite{jain_livecodebench_2024}, MATH500~\cite{hendrycks_measuring_2021}, and the Math subset of MMLU-Pro~\cite{wang_mmlupro_2024}.

Linear compensation improves LCB scores over AWQ alone at all three budgets.
The gains are 1.33--2.37 percentage points for Qwen3-4B and 1.23--1.71 points for Qwen3-8B.

\begin{table}[!t]
    \centering\footnotesize
    \caption{Reasoning scores with AWQ (\%).
    ResServe combines AWQ with linear residue compensation; BF16$^{*}$ is a reference.
    Best quantized scores per model and task are bold.}
    \label{tab:awq_reasoning}
    \setlength{\tabcolsep}{3pt}
    \begin{tabular*}{\linewidth}{@{\extracolsep{\fill}}lcrrr@{}}
        \toprule
        Method & Budget & LCB & MATH500 & \shortstack{MMLU-Pro\\Math} \\
        \midrule
        \multicolumn{5}{l}{\emph{Qwen3-4B}} \\
        BF16$^{*}$ & --- & 71.94 & 95.20 & 89.93 \\
        Plain NVFP4 & --- & 61.42 & \textbf{94.60} & 87.64 \\
        AWQ & --- & 64.55 & 94.20 & 87.86 \\
        \midrule
        ResServe & 4.5 & 65.88 & 94.00 & \textbf{89.42} \\
        ResServe & 5.0 & 66.16 & 94.40 & 88.75 \\
        ResServe & 5.5 & \textbf{66.92} & 94.40 & 89.34 \\
        \midrule
        \multicolumn{5}{l}{\emph{Qwen3-8B}} \\
        BF16$^{*}$ & --- & 73.36 & 94.60 & 92.38 \\
        Plain NVFP4 & --- & 65.59 & 95.60 & 90.97 \\
        AWQ & --- & 69.95 & 96.00 & 91.04 \\
        \midrule
        ResServe & 4.5 & 71.37 & 95.00 & 91.41 \\
        ResServe & 5.0 & \textbf{71.66} & \textbf{96.20} & 91.12 \\
        ResServe & 5.5 & 71.18 & 96.00 & \textbf{91.56} \\
        \bottomrule
    \end{tabular*}
    \vspace{12pt}
    \caption{Qwen3-8B checkpoint storage. Budgets are planner effective bits, not physical bits per weight.}
    \label{tab:budget_storage}
    \setlength{\tabcolsep}{6pt}
    \begin{tabular*}{\linewidth}{@{\extracolsep{\fill}}lrr@{}}
        \toprule
        Configuration & Size (GiB) & Increase \\
        \midrule
        Plain NVFP4 & 5.973 & --- \\
        Budget 4.5 & 6.185 & 3.55\% \\
        Budget 5.0 & 6.428 & 7.62\% \\
        Budget 5.5 & 6.542 & 9.53\% \\
        \bottomrule
    \end{tabular*}
\end{table}

\subsection{Planner Checkpoint Storage}
\label{app:planner_storage}

Table~\ref{tab:budget_storage} measures checkpoint storage in a separate set of KL-planned Qwen3-8B NVFP4 exports at the same three budgets.
The reference is an existing plain NVFP4 checkpoint.
Sizes include all files within each checkpoint directory.

The checkpoints exported at the three budgets occupy 3.55--9.53\% more storage than plain NVFP4.
Layers with partial compensation store copied weight rows, while full compensation reuses the base weight.

\begin{table}[!t]
    \centering
    \small
    \setlength{\tabcolsep}{3pt}
    \caption{System quality with W4A8 baselines (\%).
    LCB uses Hard350 pass@1; HMMT/AIME use avg@16.
    BF16 references use BF16 KV; all other rows use NVFP4 KV.
    ResServe prefill ratios are 0 for NVFP4 and 0.25 for MXFP4.
    Bold marks the highest score within each model and FP4 format.}
    \label{tab:quality-system-w4a8}
    \vskip 0.15in
    \begin{tabular*}{\linewidth}{@{\extracolsep{\fill}}llrrr@{}}
        \toprule
        \multicolumn{2}{l}{Configuration} & LCB & HMMT & AIME \\
        \midrule
        \emph{Qwen3-4B} & BF16 & 36.57 & 51.25 & 63.13 \\
        NVFP4 & Plain & 23.71 & 31.88 & 50.00 \\
        & W4A8 & 23.43 & 33.75 & 48.75 \\
        & ResServe & \textbf{28.86} & \textbf{37.92} & \textbf{58.75} \\
        MXFP4 & Plain & 10.57 & 14.58 & 28.13 \\
        & W4A8 & 19.43 & 29.17 & 48.33 \\
        & ResServe & \textbf{20.86} & \textbf{33.96} & \textbf{50.83} \\
        \midrule
        \emph{Qwen3-8B} & BF16 & 45.14 & 55.83 & 68.33 \\
        NVFP4 & Plain & 24.29 & 43.13 & 56.25 \\
        & W4A8 & 30.86 & 45.63 & 60.21 \\
        & ResServe & \textbf{35.14} & \textbf{48.96} & \textbf{61.04} \\
        MXFP4 & Plain & 12.00 & 22.08 & 36.88 \\
        & W4A8 & 21.43 & 38.33 & 51.88 \\
        & ResServe & \textbf{27.14} & \textbf{40.63} & \textbf{55.21} \\
        \midrule
        \emph{Qwen3-32B} & BF16 & 51.14 & 64.38 & 75.63 \\
        NVFP4 & Plain & 42.29 & 55.21 & \textbf{71.46} \\
        & W4A8 & \textbf{46.57} & 58.75 & 70.42 \\
        & ResServe & 46.29 & \textbf{61.67} & 71.25 \\
        MXFP4 & Plain & 31.43 & 45.83 & 57.71 \\
        & W4A8 & \textbf{40.29} & \textbf{55.00} & 64.58 \\
        & ResServe & 38.00 & 52.50 & \textbf{67.08} \\
        \midrule
        \emph{Qwen3.8-27B} & BF16 & 84.00 & 79.38 & 84.17 \\
        NVFP4 & Plain & 79.43 & \textbf{80.83} & \textbf{88.75} \\
        & W4A8 & \textbf{81.71} & 77.50 & 83.96 \\
        & ResServe & 77.71 & 78.96 & 88.33 \\
        MXFP4 & Plain & 70.86 & 66.25 & 72.92 \\
        & W4A8 & \textbf{77.71} & 72.08 & \textbf{79.79} \\
        & ResServe & 74.86 & \textbf{72.29} & 78.75 \\
        \midrule
        \emph{Swift-27B} & BF16 & 83.43 & 76.25 & 82.08 \\
        NVFP4 & Plain & 76.57 & 75.42 & 81.46 \\
        & W4A8 & \textbf{80.57} & 71.25 & 79.38 \\
        & ResServe & 77.71 & \textbf{78.33} & \textbf{85.42} \\
        MXFP4 & Plain & 65.71 & 62.50 & 67.71 \\
        & W4A8 & 75.43 & 66.67 & 74.17 \\
        & ResServe & \textbf{77.43} & \textbf{69.17} & \textbf{75.83} \\
        \bottomrule
    \end{tabular*}
\end{table}

\begin{table}[!t]
    \centering\small
    \caption{Complete checkpoint storage with uniform compensation ratios.
    Sizes and additions are in GiB; increases are relative to same-format plain checkpoints and computed from unrounded byte counts.}
    \label{tab:ratio_storage}
    \setlength{\tabcolsep}{3pt}
    \begin{tabular*}{\linewidth}{@{\extracolsep{\fill}}lrrrr@{}}
        \toprule
        Model & Plain & Size & Added & Increase \\
        \midrule
        \multicolumn{5}{l}{NVFP4, ratio 0.125} \\
        Qwen3-4B & 2.643 & 2.884 & 0.240 & 9.10\% \\
        Qwen3-8B & 5.973 & 6.431 & 0.458 & 7.67\% \\
        Qwen3-32B & 19.262 & 21.316 & 2.053 & 10.66\% \\
        Qwen3.8-27B & 19.201 & 20.801 & 1.601 & 8.34\% \\
        Swift-27B & 19.201 & 20.801 & 1.601 & 8.34\% \\
        \midrule
        \multicolumn{5}{l}{NVFP4, ratio 0.25} \\
        Qwen3-4B & 2.643 & 3.123 & 0.480 & 18.16\% \\
        Qwen3-8B & 5.973 & 6.888 & 0.915 & 15.32\% \\
        Qwen3-32B & 19.262 & 23.365 & 4.103 & 21.30\% \\
        Qwen3.8-27B & 19.201 & 22.399 & 3.198 & 16.66\% \\
        Swift-27B & 19.201 & 22.399 & 3.198 & 16.66\% \\
        \midrule
        \multicolumn{5}{l}{MXFP4, ratio 0.125} \\
        Qwen3-4B & 2.537 & 2.797 & 0.259 & 10.22\% \\
        Qwen3-8B & 5.770 & 6.204 & 0.433 & 7.51\% \\
        Qwen3-32B & 18.354 & 20.294 & 1.940 & 10.57\% \\
        Qwen3.8-27B & 18.492 & 20.005 & 1.512 & 8.18\% \\
        Swift-27B & 18.492 & 20.005 & 1.512 & 8.18\% \\
        \midrule
        \multicolumn{5}{l}{MXFP4, ratio 0.25} \\
        Qwen3-4B & 2.537 & 2.991 & 0.454 & 17.87\% \\
        Qwen3-8B & 5.770 & 6.635 & 0.865 & 14.99\% \\
        Qwen3-32B & 18.354 & 22.230 & 3.876 & 21.12\% \\
        Qwen3.8-27B & 18.492 & 21.514 & 3.021 & 16.34\% \\
        Swift-27B & 18.492 & 21.514 & 3.021 & 16.34\% \\
        \bottomrule
    \end{tabular*}
    \par\vspace{12pt}
    \centering\small
\setlength{\tabcolsep}{3pt}
\caption{NVFP4 HTTP latency (ms) across selective prefill ratios under the protocol in Appendix~\ref{app:e2e_protocol}.
Values are means across three independent server processes.}
\label{tab:e2e-nvfp4-ratios}
\begin{tabular*}{\linewidth}{@{\extracolsep{\fill}}lrrrr@{}}
    \toprule
    Method & Ratio & TTFT & TPOT & E2E latency \\
    \midrule
    \multicolumn{5}{l}{\emph{Qwen3-8B}} \\
    W4A4 & --- & 90.90 & 2.940 & 840.48 \\
    ResServe & 0 & 91.58 & 3.048 & 868.76 \\
    ResServe & 0.125 & 96.56 & 3.035 & 870.45 \\
    ResServe & 0.25 & 100.82 & 3.061 & 881.35 \\
    \midrule
    \multicolumn{5}{l}{\emph{Qwen3-32B}} \\
    W4A4 & --- & 274.27 & 8.520 & 2446.96 \\
    ResServe & 0 & 273.96 & 8.521 & 2446.91 \\
    ResServe & 0.125 & 300.80 & 8.868 & 2562.15 \\
    ResServe & 0.25 & 321.11 & 8.983 & 2611.87 \\
    \bottomrule
\end{tabular*}

\end{table}
\clearpage

\FloatBarrier
\end{document}